\documentclass[preprint,12pt]{elsarticle}

\usepackage{graphicx}

\usepackage{amsmath}
\usepackage{amssymb}

\usepackage{amsthm}

\newcommand{\rbold}{\ensuremath{\textbf{r}}}

\newcommand{\kronecker}[1]{\ensuremath{\delta_{#1}}}

\newcommand{\gabr}[2]{\ensuremath{g_{#1}(#2)}}

\newcommand{\sabq}[2]{\ensuremath{S_{#1}(#2)}}

\journal{Chemical Physics Letters}

\begin{document}

\begin{frontmatter}



\title{Radical re-appraisal of water structure in hydrophilic confinement}


\author{Alan K. Soper}

\address{ISIS Facility, STFC Rutherford Appleton Laboratory, Harwell Oxford, Didcot, OX11 0QX, UK}

\begin{abstract}
The structure of water confined in MCM41 silica cylindrical pores is studied to determine if confined water really is simply a version of the bulk liquid which can be substantially supercooled without crystallisation. A combination of total neutron scattering from the porous silica, both wet and dry, and computer simulation using a realistic model of the scattering substrate is used. The water in the pore is divided into three regions: core, interfacial and overlap. The average local densities of water in these simulations are found to be about 20\%\ lower than bulk water density, while the density in the core region is below, but closer to, the bulk density. There is a decrease in both local and core densities when the temperature is lowered from 298K to 210K. The radical proposal is made here that water in hydrophilic confinement is under significant tension, around -100MPa, inside the pore.

\end{abstract}


\end{frontmatter}


\section{Introduction}
\label{intro}

Does water have a second (liquid-liquid) critical point? The answer to this bold but vexing proposal, which first came to prominence over 20 years ago \cite{poole1992phase}, is neither simple nor free from controversy. A host of computer simulations \cite{poole1992phase, stanley1994there,harrington1997liquid,brovchenko2005liquid,liu2009low,abascal2010widom,bertrand2011peculiar,liu2012liquid} and theories \cite{poole1993phase, poole1994effect,ponyatovskii1994second,sastry1996singularity,truskett1999single,tanaka2000thermodynamic, franzese2002liquid, debenedetti2003supercooled,gibson2006metastable} have served to illustrate the ideas behind the second critical point scenario, while a series of experimental investigations \cite{mishima1998relationship,faraone2004fragile,savedrecs2005:1,liu2006quasielastic, mallamace2006fragile,chen2007:4,mallamace2007evidence,chen2009evidence,huang2009inhomogeneous} have sought to place the proposal on a practical footing. Moreover reports of a first-order-like, reversible, transition between low density amorphous ice (LDA) and equilibrated high density amorphous ice (e-HDA) \cite{nelmes2006annealed,winkel2008water,winkel2011equilibrated}, and annecdotal reports of distinct highly viscous liquids immediately above the glass transition of each of these materials, lend weight to the idea that there \textit{may} be a (hidden) second critical point somewhere at higher temperatures. Yet the fact remains that no-one has yet witnessed this second critical point in real, experimental, water, which, if it occurs at all, is positioned at a point in the phase diagram where the natural state of water is firmly as crystalline ice, rather than any form of the liquid.

Visiting Martian aliens might well be puzzled by the extensive discourse on this topic. Coming from a planet which is arguably suffering the worst drought in its 4.5 billion year history they would gaze longingly at our plentiful oceans and wonder why we talk about water “anomalies” when water is by far the most plentiful liquid on the surface of planet Earth. Surely it is water that is “normal” and other fluids, like argon or nitrogen, which are “anomalous”? Of course the knowledgable water expert will kindly explain the distinction between “simple” fluids like argon or nitrogen, and “complex” fluids like water. In the case of argon or nitrogen the structure and properties are determined rather accurately by short-ranged, mostly pairwise additive, forces between the atoms of such liquids, with many-body forces playing only a minor role \cite{teitsma1980three}. With water however the situation is far less clear. Certainly even before and since the beginning of computer simulation, water has been envisaged and \textit{simulated} with pairwise additive potentials \cite{bernal1933theory,barker1969structure,rahman1971molecular,berendsen1981interaction,jorgensen1981quantum} which involve appropriate Coulomb interactions placed on or near the atoms of the water molecule, but whether these are the correct way to represent real water remains an open question. In the past 4 years or so Molinero and coworkers \cite{molinero2008water,moore2009growing,moore2010ice, moore2011structural,xu2011there} have been performing some remarkable large scale simulations of water and water mixtures, using a short range water potential (called mW) which has no electrostatic parameters, but which does have an important 3-body, albeit still short range, term which controls the spatial distribution of neighbouring water molecules and so helps to form the random network of the liquid. (The form of this potential energy function was originally developed by Stillinger and Weber with reference to liquid and amorphous silicon \cite{stillinger1985computer}.) Tuning the strength of this three-body interaction one is apparently able to simulate a variety of tetrahedral liquids including silicon, germanium and water. For water this simple potential is able to reproduce surprisingly accurately the more important thermodynamic and structural trends of water \cite{molinero2008water}.

Yet mW is still far from being a perfect water potential. The calculated diffusion constant for this potential is a factor of ~3 too large compared to experimental water \cite{molinero2008water}, and this might be one reason that a liquid-liquid critical point is not observed with this potential: at the relevant temperatures, ice crystallisation proceeds at a faster rate than that needed to equilibriate the two liquids below the critical point, which is of course presumably what happens in the real liquid, and so prevents any possible observation of an actual critical point. The original determination of the water 2nd critical point was based on the ST2 potential of Rahman and Stillinger, but other common water potentials either give this second critical point at a different point in the water phase diagram, or, like mW water, do not show a second critical point at all. Some authorities even claim that none of the common water potentials show a second critical point \cite{limmer2011putative,limmer2012phase,limmer2013putative}. Hence there is still much debate and uncertainty about the existence of a second critical point even in simulated water.

Is there anything we can learn from experiment about the nature of water in the supercooled regime? As with the theoretical understanding, the challenges for the supercooled water experiment are substantial and may be prohibitive. Water of course readily crystallises below 273K and this crystallisation can be instigated by the tiniest amounts of impurities. Below about 235K crystallisation proceeds spontaneously without the need for impurities. The only way, apparently, to avoid this \textit{homogeneous} crystallisation is by confining water in a matrix, either a liquid matrix such as an emulsion \cite{bellissent1989structural}, or in a porous solid substrate, such as in Vycor glass \cite{thompson2007three}, or the MCM glasses \cite{schreiber2001melting}. None of these methods for avoiding crystallisation is ideal however since there then arises the question of to what extent the confinement is affecting the properties of the water. Can this confined water be correctly regarded simply as bulk water for which the freezing transition has been inhibited, or is real confined water more complicated than this simple view? Once again opinions differ widely on this significant question, with support for either view, and certainly little consensus. The problem here is quite analogous to that of the behaviour of an animal in a cage compared to its behaviour in its natural environment. The animal is the same in both cases, but is its behaviour the same?

As the title suggests this paper is devoted to trying to determine and understand one aspect of this problem, namely the structure of water in confinement and how it compares with bulk water structure. Are the two structures the same or renormalisable in some sense, or are there some more fundamental differences that preclude the possibility of relating the properties of confined water to their bulk liquid counterparts? To answer these questions, we first need to understand how you would measure the structure of a liquid in the bulk form, then see how these methods need to be modified to deal with the confined liquid. Fortunately measuring the structure of bulk water has been the subject of recent extensive and independent reviews \cite{skinner2013benchmark,soper2013radial}, which give excellent agreement with each other, and the underlying computer simulation methodology used to interpret the experimental data has also been given recent expositions \cite{soper2012computer,soper2013empirical}. Hence the bulk of this article can concentrate on the modifications to these techniques needed to study confined water. We will focus on the particular case of water in the porous silica, MCM41, which (in principle) consists of long, parallel cylindrical pores arranged on a simple hexagonal lattice, with the substrate generally believed to be amorphous or partly crystalline, depending on the exact method of preparation. This does not preclude other possible systems, such as clay systems, but simply reflects the fact that this system has been widely studied, that good quality scattering data from water confined in MCM41 is available \cite{mancinelli2009multiscale}, and that the underlying structure of the substrate is sufficiently well defined that realistic computer simulation models of its structure can be built.

\section{Measuring the structure of a liquid}
\subsection{\label{theory}Theory}

Because it lacks long range order, the atomic-scale structure of any disordered material is characterised by measuring or calculating the correlations of one atom or molecule with respect to another. The simplest correlation function is the pair correlation function, which, as its name implies, measures the correlations between pairs of atoms. Three-body and higher order correlation functions can be defined and may be important in particular cases, but the pair correlation function is the simplest to define, and moreover makes direct contact with the radiation scattering properties of the material. Given a beam of radiation (x-rays, electron, neutrons) of wavelength $\lambda$ scattered by a material by angle $2\theta$, the scattered intensity as a function of $Q=\frac {4\pi\sin\theta} {\lambda}$ is given by:-

\begin{equation}
F(Q)=\sum_{\alpha}c_{\alpha}\langle b_{\alpha}^{2}\rangle + \sum_{\alpha\beta\ge\alpha}(2-\kronecker{\alpha\beta})c_{\alpha}c_{\beta}\langle b_{\alpha}\rangle \langle b_{\beta}\rangle\sabq{\alpha\beta}{Q}
\label{eq2l}
\end{equation}
with the partial structure factors defined by
\begin{eqnarray}
\sabq{\alpha\beta}{Q}&=&4\pi\rho\int r^{2}(\gabr{\alpha\beta}{r}-1)\frac{\sin Qr}{Qr}dr
\label{eq2l1}
\end{eqnarray}
for an isotropic system. Here \gabr{\alpha\beta}{r} is the set of site-site radial distribution functions that will be used to define the structure of the fluid, although it is important to remember, for the later discussion about water in confinement, that these functions are themselves obtained from the auto-correlation of the single particle density functions, $N_{\alpha}(\rbold)$: for a bulk fluid $\langle N_{\alpha}(\rbold)\rangle$ will be uniform, but in confinement $\langle N_{\alpha}(\rbold)\rangle$ will vary with displacement \rbold, even after ensemble averaging.

In equation (\ref{eq2l}) $b_{\alpha}$ represents the scattering length of the atom $\alpha$. For neutrons this is simply a number which depends on the spin and isotope states of the atomic nucleus, so the angle brackets represent averages over these spin and isotope states of the respective nuclei. These averages are not needed for electrons and x-rays where the scattering lengths are called ``form factors'' which are $Q$ dependent and which depends on the electron distribution in the atom. For the present work we only consider neutron scattering data, and will exploit the fact that hydrogen atoms have a different neutron scattering length (-3.74fm) compared to deuterons (6.67fm) \cite{searstable1992}. This contrast means that in the case of water, according to (\ref{eq2l}), experiments on different samples of the same material where some or all of the protons have been replaced with deuterons can in principle be used to extract the three site-site radial distribution functions for water, namely \gabr{OO}{r}, \gabr{OH}{r} and \gabr{HH}{r} (strictly the corresponding partial structure factors, \sabq{OO}{Q}, \sabq{OH}{Q} and \sabq{HH}{Q}, then numerically inverting the Fourier transform (\ref{eq2l1})). 

Nowadays this potentially error-prone process has been replaced by an alternative approach which involves running a computer simulation of the material in question, then refining the empirical potential used in that simulation to give the best possible agreement with the measured data \cite{soper2001tests,soper2005partial,soper2010network,soper2012computer,soper2013radial}. For systems which contain more than three components, such as the case of water confined in MCM41 discussed here, or where suitable isotopic contrasts are not available, extracting site-site distribution functions from the scattering data is not possible even in principle, so in order to understand what the data are telling us structurally, there is little alternative but to run a computer simulation to assist in the process of understanding the measurements. Of course, given that there will normally be fewer datasets than the number of site-site distribution functions required to define the structure in such cases, any computer simulated reconstruction of the real material may be prone to ambiguities and uncertainties. However the computer simulation approach allows the introduction of known constraints, such as limits on the nearest-neighbour approach, occurrence of hydrogen bonds, and, in the present case in particular, the hexagonal porous nature of the substrate, which can help to reduce the uncertainties from lack of measured information. For the case of confined water discussed here a simplified hexagonal arrangement of cylindrical pores in an amorphous silica matrix will be used as the starting point for these simulations.

Besides the structurally important scattering, as defined by the second term in equation (\ref{eq2l}), the scattering data contain an additional term, the so-called ``single atom scattering'', which is given by the first term in (\ref{eq2l}), that is $\sum_{\alpha}c_{\alpha}\langle b_{\alpha}^{2}\rangle$. This terms arises from the diagonal components of the scattering matrix, and represents the correlation of each atom with itself. It contains no structural information, but represents the scattering level about which the structural correlations oscillate. Because this is a known quantity for each material being investigated (provided the composition of the material is known), the single atom scattering provides a simple level to determine the absolute normalisation of the data. This is particularly important when, as at present, the scattering sample occurs in a powdered form which does not fill the sample containment completely. There is in such cases an unknown ``packing fraction'' which has to be determined in order to put the scattering data onto an absolute scale of differential cross section. The single atom scattering level can be used to do exactly that.

With hydrogen-deuterium substitution there is a slight complication to this process, which arises from the fact that the spin averaged single atom scattering from a proton (H) is more than 10 times larger than than from a deuteron (D) \cite{searstable1992}. Because of the very ready exchange of H for D when heavy water is exposed to the atmosphere, even a small amount of H present in an ostensibly fully deuterated sample can have a marked impact on the scattering level. Hence uncertainties about the exact amount of H present can mar our ability to put the scattering data onto an absolute scale of differential scattering cross section. The basic steps used to reduce raw scattering data to differential scattering cross section are described in several places, e.g. \cite{fischer2006neutron}, and the particular methods used here are given in \cite{soper2011gudrunn}. The actual preparation of the scattering data used for the subsequent structure refinement has already been described in detail by Mancinelli \textit{et al.} \cite{mancinelli2009multiscale} and so will not be repeated here. However it is worth pointing out that the MCM41 materials used in these experiments derived from the same source as those used by Liu et al. \cite{liu2007observation} In addition, for the heavy water samples, it is assumed, for both the data analysis and subsequent interpretation of the data using computer simulation, that the D$_{2}$O was contaminated with 10\%\ H$_{2}$O. There is no actual evidence for such contamination, although it is not impossible given the amount of sample handling that is involved in these experiments. However it enabled the amount of water inserted into the simulation box to approach that deduced from the experiment (0.43g H$_{2}$O per g substrate) and still retain an acceptable fit to the data.

\subsection{\label{epsr}Emprical potential structure refinement}

\subsubsection{\label{drymcm}Simulations of dry MCM41}

The method used here to model the scattering data is a development of the empirical potential structure refinement (EPSR) method that has been described in several recent publications \cite{soper2001tests,soper2005partial,soper2010network,soper2012computer,soper2013radial}. EPSR was itself derived from the Reverse Monte Carlo (RMC) method \cite{mcgreevy1988reverse}. Although similar in scope, EPSR is distinguished from RMC by using the difference between scattering data and simulated structure factors to develop a perturbation to an initial seed potential, called the ``reference'' potential. This perturbation is called the ``empirical'' potential and it aims to drive the simulated structure factors as close as possible to the measured data.

To build the initial model of (dry) MCM41, a line of 80 ``pseudo'' atoms, called $q$-atoms, spaced 1.85\AA\ apart, are placed at the centre of a hexagonal unit cell ($a,b,c$) of dimensions (33.1\AA, 33.1\AA, 148.0\AA), with an angle of 120$^\circ$ between the crystallographic $a,b$ axes. This line is parallel to the crystallographic $c$-axis, which in turn is perpendicular to the $a,b$ plane. The unit cell is repeated once along each of the $a$ and $b$ axes to give a 2$\times$2 supercell. The $q$-atoms are given a diameter of 25\AA, which was determined from an earlier analysis, \cite{soper2012density}, and 5420 silicon and 10840 oxygen atoms are inserted at random into the available space around the $q$-atoms, this number corresponding to approximately 90\%\ of the atomic number density of bulk silica, namely $\sim$0.066 atoms/\AA$^3$. To prevent these atoms moving into the pore during the subsequent computer simulation, a repulsive potential of the form $U_{\alpha\beta}^{(rep)}(r)=C_{\alpha\beta}\exp\left[\frac{1}{\gamma}\left(r_{\alpha\beta}-r\right)\right]$, where $r_{\alpha\beta}$ and $\gamma$ are the specified minimum distance for atom pair $\alpha,\beta$ and ``hardness'' parameter respectively, is applied between the $q$-atoms and the silicon and oxygen atoms. As described in \cite{soper2013radial}, the amplitude $C_{\alpha\beta}$ is adjusted automatically as the simulation proceeds, depending on the extent to which atoms of the respective pair of atom types are found below the specified minimum distance. The full set of parameters for the atoms and molecules used in these simulations is given in Table \ref{potpars} and it will be seen that the $r_{qSi}$ values are larger than $r_{qO}$, so allowing the oxygen atoms to penetrate slightly further into the pores than the silicon atoms.

\begin{figure}
\begin{tabular}{c}
(a)  \\
\includegraphics[width=1\textwidth]{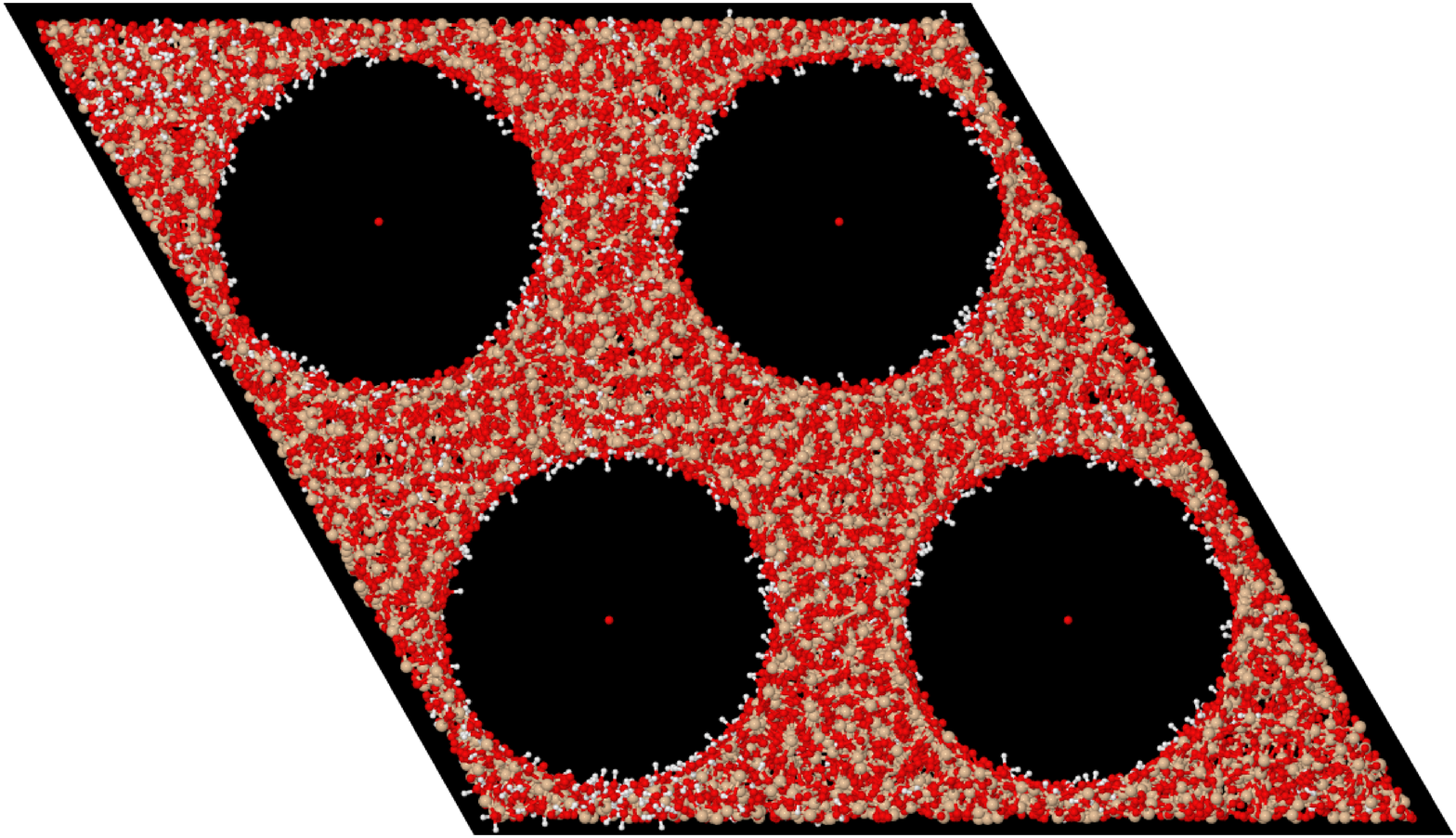} \\
(b) \\
\includegraphics[width=1\textwidth]{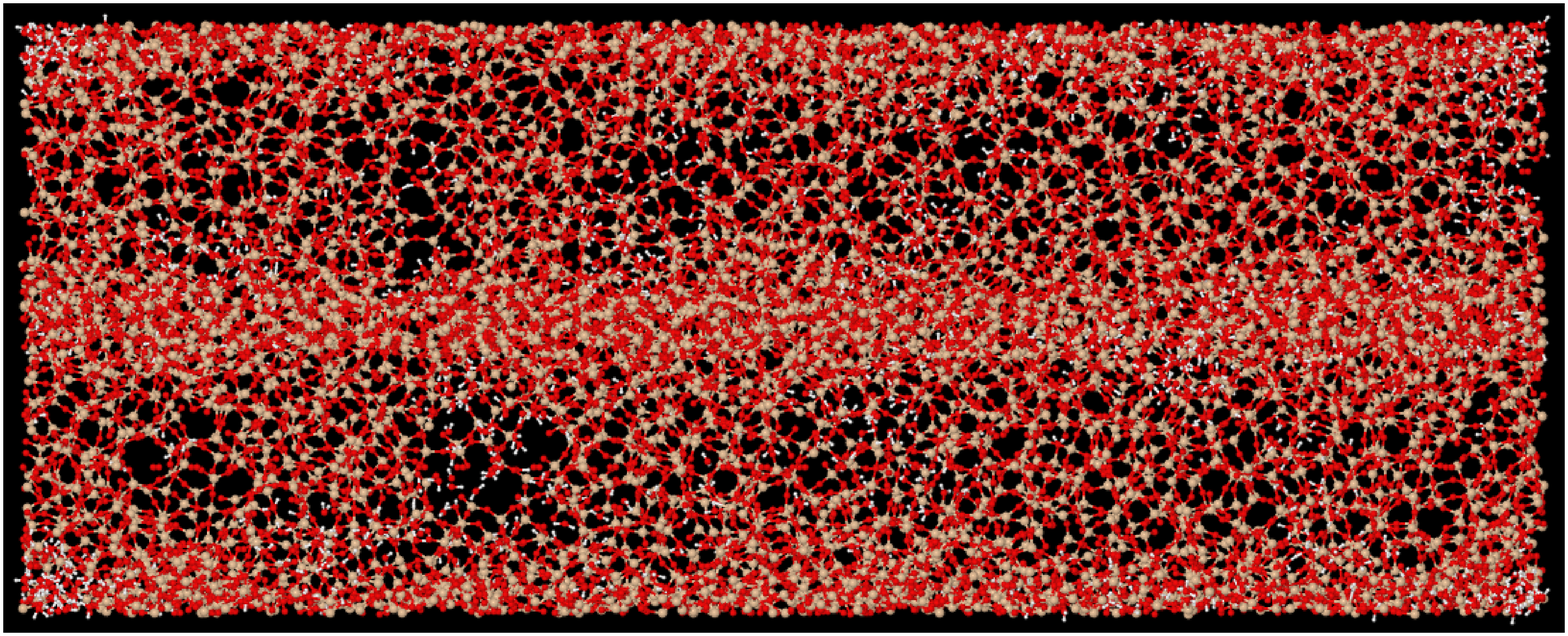}
\end{tabular}
\caption{(Colour online) Computer simulation box of dry MCM41, along the $c$-axis (a) and at right angles to the $c$-axis (along the $a$-axis (b). The dimensions of the box are 66.2\AA\ along each of the $a$ and $b$ axes, and 148\AA\ along the $c$ axis. The small red dots at the centre of each pore represent the $Q$-atoms mentioned in the text: these make no contribution to the scattering pattern, but are used simply to prevent silica and ``silanol'' water molecules from entering the pores. Silanol water molecules populate the surface of the pores, but some of these are seen to permeate the silica matrix as well.}
\label{figdrymcm41} 
\end{figure}

\begin{figure}
\centering
\includegraphics[scale=0.7]{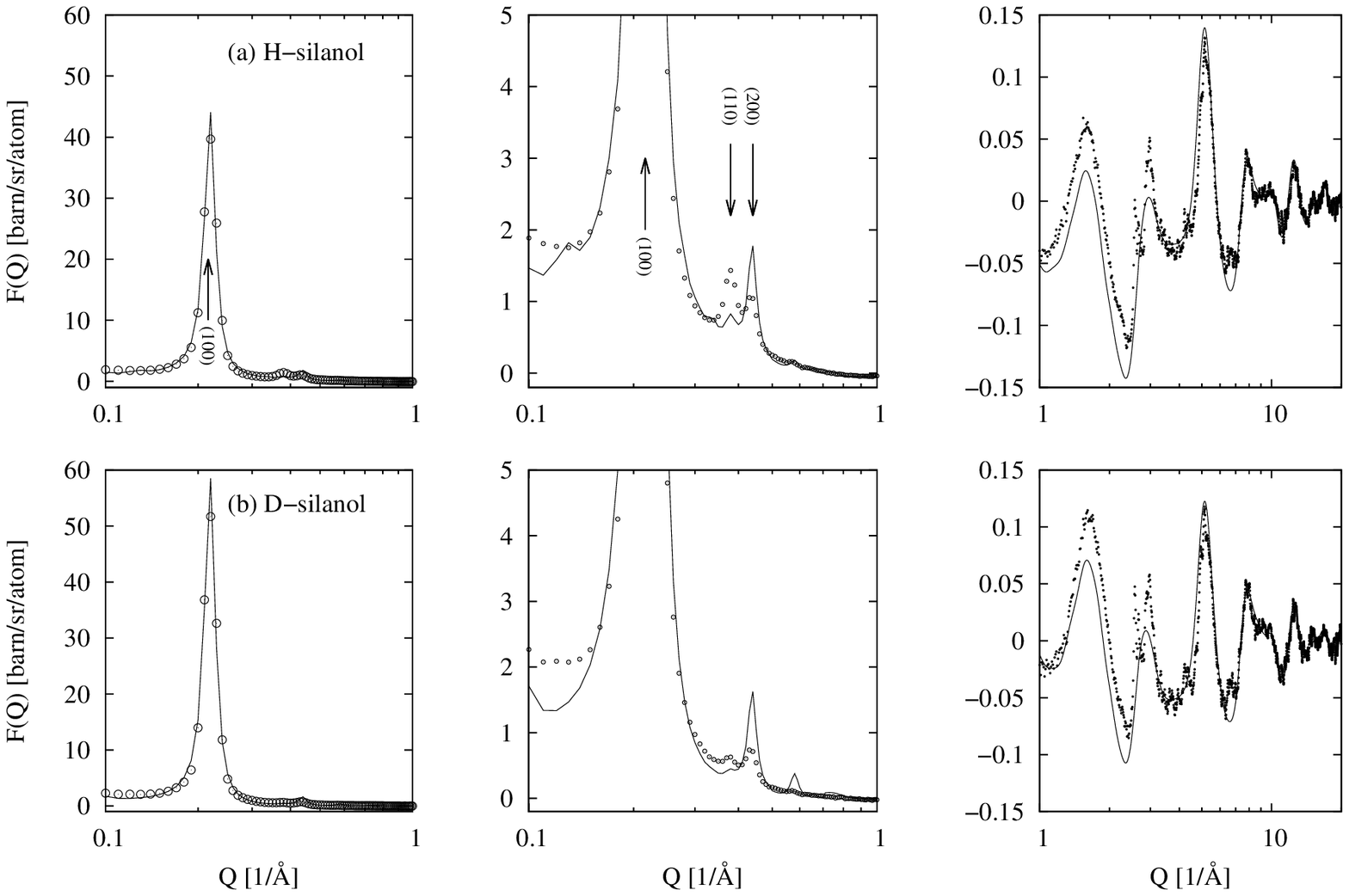}
\caption{EPSR fit (line) to the total scattering data from dry MCM41 (circles) over three different scales of intensity and $Q$. Row (a) shows the fits to the data with protiated silanol water molecules, while row (b) shows the fit to the data with deuteriated silanol water molecules.The left-hand plot shows the (100) Bragg peak, the middle plot shows the (110), (200) and higher order Bragg peaks, while the right-hand plot shows the wider $Q$ region beyond $Q$ = 1\AA$^{-1}$. }
\label{figdrymcm41fits}
\end{figure}

To emulate the silanol groups that invariably populate the surface of these MCM41 pores, a number of water molecules are also introduced into the silica matrix, at the rate of 0.179 water molecules per silicon atom, this proportion having been determined in previous work \cite{mancinelli2009multiscale}. The atoms of these ``silanol'' water molecules are given the labels OS and HS respectively to distinguish them from the corresponding water molecule atoms, OW and HW, which will be introduced into the pore. The silanol water molecules are also constrained not to enter the pore - see Table \ref{potpars} - but are otherwise not prevented from entering the silica matrix if required. The use of water molecules to represent the silanol groups in this way preserves electrical neutrality while also maintaining the required stoichiometry. The OS atoms of these silanol water molecules have identical Lennard-Jones and Coulomb potential parameters to those of the silica oxygen atoms, so can in principle substitute for those atoms as needed. Calculation of the (100), (110) and (200) Bragg peaks from the hexagonal lattice was performed directly on the simulation box using the methods described in \cite{soper2013empirical}. Figure \ref{figdrymcm41} gives a snapshot of the simulation box after structure refinement against the dry MCM41 data, while Figure \ref{figdrymcm41fits} shows the fit to the total scattering data for both protiated and deuteriated materials.

The EPSR model captures the different amplitudes of the (100) Bragg peak (Fig. \ref{figdrymcm41fits}, left panels) for the two samples (protiated and deuteriated silanol groups) quite accurately, but less accurately for the higher order Bragg peaks (Fig. \ref{figdrymcm41fits}, middle panels), although the very rapid decline in intensity in these peaks is captured qualitatively. At higher $Q$ values (Fig. \ref{figdrymcm41fits}, right panels) the underlying structure is captured mostly quantitatively by the EPSR model with some discrepancies in peak heights. Improving on these fits is in principle possible, but it has to be recorded that the calculated intensities from the EPSR model here are in absolute units differential cross section, while the normalisation of the scattering data is achieved by ensuring the single atom scattering level in the scattering data (first term in equation \ref{eq2l1}) is consistent with the stated atomic composition. The EPSR model assumes a perfect crystal but almost certainly the real material has significant defects, so that insisting on too good a fit could generate spurious structure. In particular there is a degree of local crystallinity in this substrate, as witnessed by the sharp peaks in the total scattering data, Fig. \ref{figdrymcm41fits}, right panels, which the present EPSR method will never capture without more detailed atom-scale information becoming available.

Once the equilibrium in this simulation had been reached and the fit to the scattering data could not be improved further, the silica atoms and silanol water molecules were ``tethered'' to their current positions. This means in subsequent simulation steps the atoms can move around these positions by small amounts, but cannot diffuse away. This refined simulation box was then used as the substrate on which to absorb water molecules into the pores. Note that for all the simulations reported in this work, the range of both the reference and empirical potentials was set to 30\AA: this was needed to ensure the simulation captured the longer range correlations implicit in the scattering data.

\begin{table}
\begin{center}
\begin{tabular}{|c|ccc|}
\hline 
\multicolumn{4}{|c|}{Lennard-Jones parameters and Coulomb charges} \\ 
\hline 
Atom & $\epsilon$ & $\sigma$ & $q$ \\ 
• & [kJ/mol] & [\AA] & [e] \\ 
\hline 
$q$ & 0.00 & 0.00 & +0.0000 \\ 
Si & 0.80 & 1.06 & +2.0000 \\ 
O & 0.65 & 3.09 & -1.0000 \\ 
OS & 0.65 & 3.09 & -1.0000 \\ 
HS & 0.00 & 0.00 & +0.5000 \\ 
OW & 0.65 & 3.16 & -0.8476 \\ 
HW & 0.00 & 0.00 & +0.4238 \\ 
\hline 
\multicolumn{4}{|c|}{Minimum distances} \\
\hline 
Atom pair & \multicolumn{3}{c|}{$r_{\alpha\beta}$} \\ 
& \multicolumn{3}{c|}{[\AA]} \\ 
\hline 
$q$-Si & \multicolumn{3}{c|}{12.0} \\ 
$q$-O & \multicolumn{3}{c|}{11.5} \\ 
$q$-OS & \multicolumn{3}{c|}{11.5} \\ 
$q$-HS & \multicolumn{3}{c|}{10.5} \\
Si-OW & \multicolumn{3}{c|}{2.50} \\ 
\hline
\end{tabular}
\end{center} 
\caption{\label{potpars}Lennard-Jones and Coulomb parameters (top) and minimum distances (bottom) for the reference potentials used in the EPSR simulations described in this work.} 
\end{table}
 
\subsubsection{\label{wetmcm41}Simulations of wet MCM41}

In the experiment, water was allowed to enter the MCM41 matrix at the rate of $\approx$0.43g/g of substrate, \citep{mancinelli2009multiscale}. This was measured by weighing the sample before exposure to water vapour and after exposure. Absorbing water in this way helped to ensure as much water as possible is absorbed inside the pore, although the exact amount in the pore as opposed the external surface of the silica particles is difficult to ascertain precisely. Only very weak ice Ih Bragg peaks appeared when the sample was cooled below 273K, suggesting the amount of surface water present was small.

In setting up the EPSR simulation of wet MCM41 it became apparent that allowing as much water as would be implied by the experimental 0.43g/g into the pores would make fitting the data, particularly those from the deuteriated samples, difficult. As will be seen shortly, there is approximately a factor of 4 reduction of the (100) Bragg peak intensity going from H$_2$O absorbed samples to D$_2$O samples, and if the amount of water entered into the pore is too large, this intensity ratio is difficult to reproduce. After some experimentation with the amount of water in the pore, the amount used in the simulations presented here corresponds to 0.39g (H$_2$O) per g (SiO$_2$), giving, if the silanol water molecules are included, a composition close to, but slightly below, the experimental value. The total number of water molecules added to the four pores of Fig. \ref{figdrymcm41} was 7046, these being initially distributed randomly within the confines of the pore. However, and unlike the silanol water molecules, a mild restriction on pore water entering the silica matrix was imposed. This was achieved by specifying a minimum separation of 2.5\AA\ on the Si-OW interactions. This did not exclude water completely from the silica matrix, but increasing this minimum separation forced water out of the silica matrix to an increasing extent. Hence this minimum separation was used as a control on how much pore water could penetrate the silica, and its value was chosen so that after structure refinement the ratio of simulated (100) Bragg peak intensities between H$_2$O and D$_2$O matched that observed in the experiment. The same minimum separation and numbers of water molecules were used for simulations at both 298K and 210K, where the only differences were the temperatures of the simulations and the scattering data against which they were refined. Table \ref{potpars} also lists the Lennard-Jones and charge parameters, based on the SPC/E water potential \cite{berendsen1987missing}, used as the reference potential for these molecules.

\begin{figure}
\centering
\includegraphics[scale=0.7]{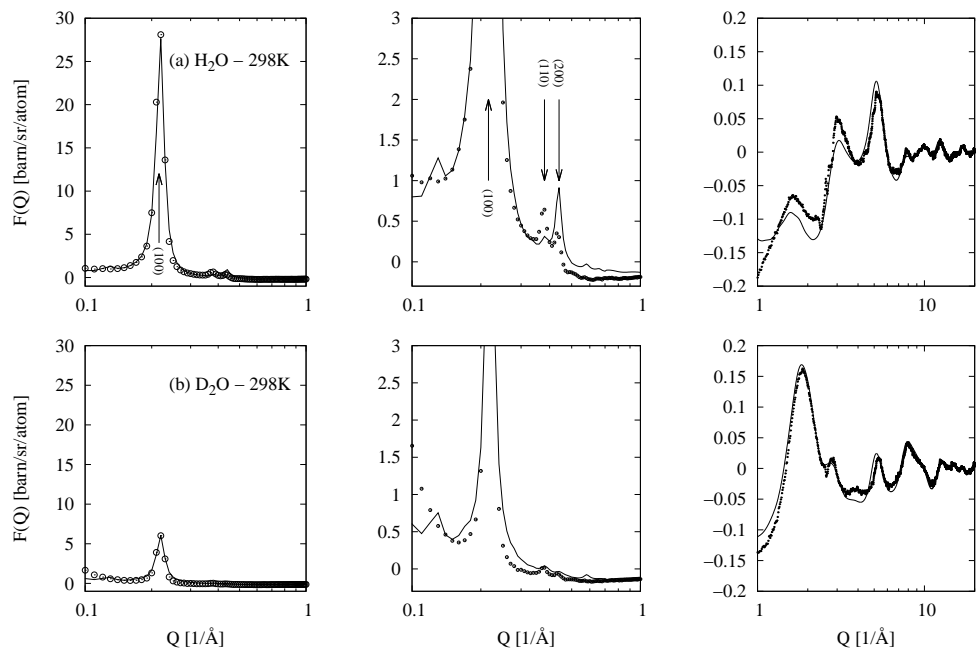}
\caption{EPSR fit (line) to the total scattering data from wet MCM41 at 298K (circles) over three different scales of intensity and $Q$. Row (a) shows the fits to the data with absorbed H$_2$O water molecules, while row (b) shows the fit to the data with D$_2$O water molecules.The left-hand plot shows the (100) Bragg peak, the middle plot shows the (110), (200) and higher order Bragg peaks, while the right-hand plot shows the wider $Q$ region beyond $Q$ = 1\AA$^{-1}$. }
\label{figwetmcm41fits}
\end{figure}

\begin{figure}
\centering
\includegraphics[scale=0.7]{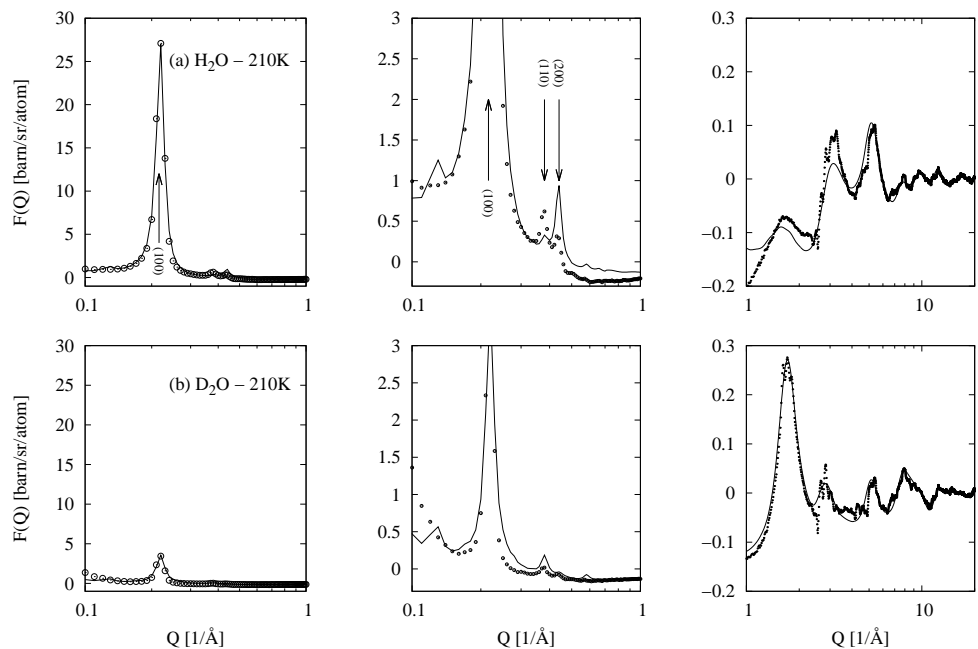}
\caption{EPSR fit (line) to the total scattering data from wet MCM41 at 210K (circles) over three different scales of intensity and $Q$. Row (a) shows the fits to the data with absorbed H$_2$O water molecules, while row (b) shows the fit to the data with D$_2$O water molecules.The left-hand plot shows the (100) Bragg peak, the middle plot shows the (110), (200) and higher order Bragg peaks, while the right-hand plot shows the wider $Q$ region beyond $Q$ = 1\AA$^{-1}$. }
\label{figwetmcm41fits210K}
\end{figure}

Figure \ref{figwetmcm41fits} shows the EPSR fits to the data at 298K while figure \ref{figwetmcm41fits210K} shows the fits to the data at 210K. It should be emphasized here that both fits were obtained with the same silica substrate and the same number of water molecules in the simulation box: it was not necessary to reduce the number of water molecules at 210K compared to 298K, as has been suggested in recent publications \cite{kamitakahara2012temperature}. This matter will be discussed further in the Results and Discussion sections. A case in point is the height of the (100) Bragg peak with absorbed D$_2$O: this peak falls in intensity on lowering the temperature, a trend which is captured quite accurately with the present simulations at constant water mass. As with the dry MCM41, it was not possible to capture every detail of the data with these simulations, but the main trends in terms of peak heights and positions are reproduced correctly. There is a notable increase in the number of small Bragg peaks in the data at 210K at wider $Q$ values compared to 298K, and these cannot be reproduced by the present simulations, which assume an amorphous model for both silica substrate and water. Since these simulations are performed in a Monte Carlo framework, they also tell us nothing about the dynamics of the absorbed water molecules.

\section{\label{results}Results}

\subsection{\label{poreradius}Choice of pore radius}

The pore radius used at the outset in these simulations is 12.5\AA. This choice of value was dictated by previous considerations based on the amount of water apparently absorbed inside the pore, and also by choosing a radius that simultaneously gave an adequate fit to the data for both deuteriated and protiated silanol groups for the dry MCM41 \cite{soper2011density,soper2012density}. In practice, when a realistic, disordered model of the surface is built \cite{mahadevan2008dissociative,xu2009thermal} penetration of water into the surface can occur to a depth of several \AA, making precise characterisation of the pore diameter problematic. With silica this happens, with the added complication of silanol formation at (or near) the surface.

\begin{figure}
\centering
\includegraphics[scale=0.7]{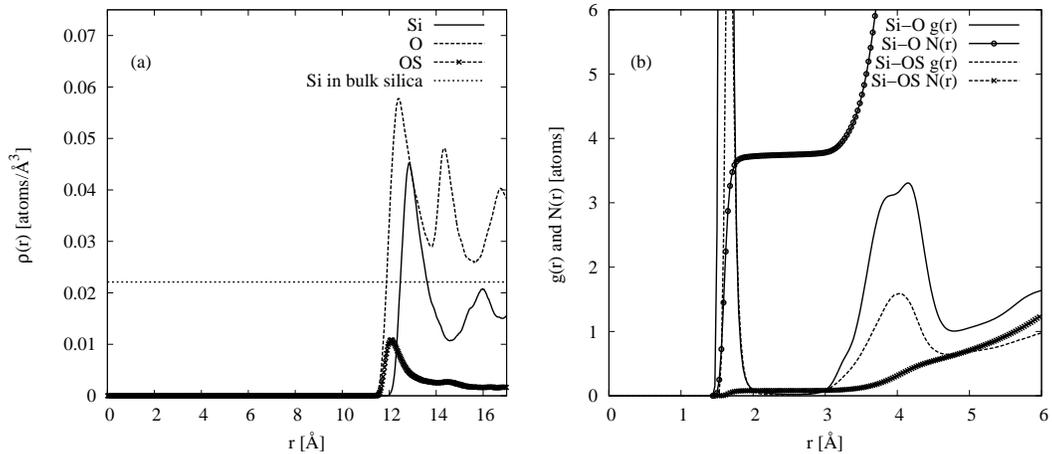}
\caption{(a) Density profile of Si (solid), O (dashed) and OS (dashed, crosses) of the simulated dry MCM41 substrate. Also shown is the expected bulk density of Si atoms (dotted). (b) Si-O (solid) and Si-OS (dashed) radial distribution functions, together with the corresponding running coordination numbers, $N_{SiO}(r)$ (solid, circles) and $N_{SiOS}(r)$ (dashed, crosses).  }
\label{figdrymcm41profiles}
\end{figure}

In the present case the final pore diameter after structure refinement was set by the choice of minimum distances between $q$ atoms and the Si and O atoms of the substrate as given in Table \ref{potpars}. To determine the radius that finally emerged using these values, Fig. \ref{figdrymcm41profiles}(a) shows the density profile of the Si, O and OS atoms as a function of distance from the pore centre. It can be seen that the pore radius achieved after structure refinement is $\sim$ 12.0\AA, based on the oxygen atom density distribution. However the silica substrate is highly structured near the surface, with a pronounced shell of silicon just below the surface, and surface oxygen (O) and silanol oxygen (OS) atoms attached to this. Fig. \ref{figdrymcm41profiles}(b) shows a detail of the Si-O and Si-OS radial distribution functions together with the corresponding Si-O and Si-OS running coordination numbers, $N_{\alpha\beta}(r)$, where
\begin{equation}
 N_{\alpha\beta}(r)=4\pi\rho_{\beta}\int_{0}^{r}r'^{2}g_{\alpha\beta}(r')dr'.
 \label{nofr}
\end{equation}
These indicate that the total oxygen coordination number in the first shell of silicon is close to 4, as happens in bulk amorphous silica.

The mean density of the silicon atoms in the substrate appears lower than that in the bulk density, Fig. \ref{figdrymcm41profiles}(a). However attempts to increase this amount to get closer to the bulk density gave poorer fits to the scattering data. In section \ref{localdensity} below we introduce the concept of a \textit{local} density for heterogeneous systems and discuss this matter in more detail.

\subsection{\label{rdfs}Radial distribution functions}

\begin{figure}
\centering
\includegraphics[scale=0.7]{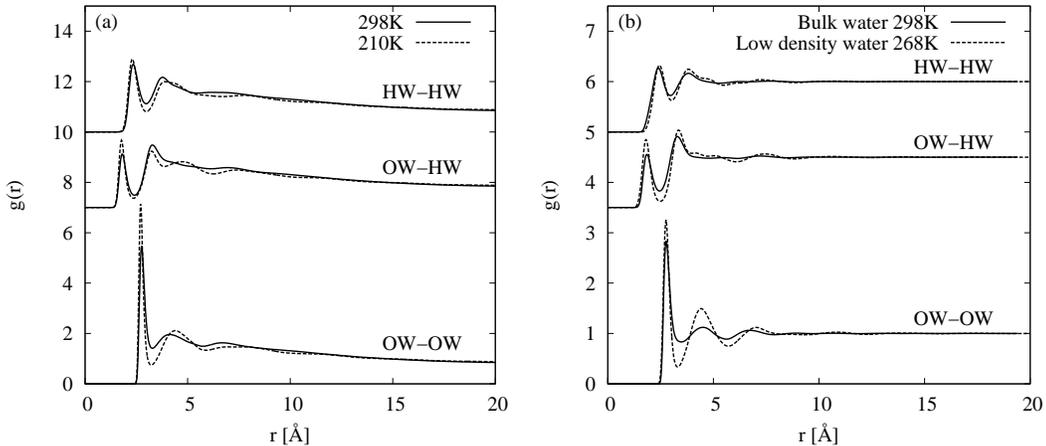}
\caption{(a) OW-OW, OW-HW and HW-HW radial distribution functions for water absorbed in MCM41 at 298K (solid) and at 210K (dashed) (b) Same functions for bulk water (solid) \cite{soper2013radial}. Also shown are the estimated radial distribution functions for low density water at 268K (dashed), as derived in \cite{soper2000structures}. In both cases the OW-HW and HW-HW are shifted vertically for clarity.}
\label{figwaterrdfs}
\end{figure}

The simulated radial distribution functions for water in MCM41 are shown in Fig. \ref{figwaterrdfs}(a). Comparing these with those found in bulk water \cite{soper2013radial}, using the same SPC/E reference potential, Fig. \ref{figwaterrdfs}(b), some marked differences can be seen. 

Firstly in confined water the data are on a marked negative slope with increasing $r$ and only reach $g(r)=1$ at $r > 30$\AA, while the bulk data oscillate about $g(r) = 1$ for all $r$. At the same time the amplitude of the peaks is roughly a factor of 2 larger for water in confinement compared to their bulk counterparts. These effects are well-known from other studies of confined fluids \cite{lee1992local,thompson2007three, mancinelli2010effect} and are labelled as ``excluded volume'' effects, arising as they do from the fact that the fluid is excluded from some regions of the sample \cite{soper1997excluded}. However they make direct comparison with the bulk fluid difficult, Fig. \ref{figwaterrdfs}(b), unless one is prepared to develop a fairly elaborate correction procedure which takes account of the density variation both inside and outside the pore \cite{thompson2007three,mancinelli2009multiscale,mancinelli2010effect}. 

At low $r$ the confined water distributions oscillate about a level of $\sim 2$, suggesting that the local density of the water in the pore is roughly twice the density of water averaged over the full volume of the MCM41 unit cell. Taking account of this local density effect, the first two peaks in the water radial distribution functions are about the same heights and positions as their bulk water counterparts. There may be a slight distortion towards lower $r$ for the second peak in the OW-OW function, as was seen by \citep{mancinelli2009multiscale,mancinelli2010effect}.

On cooling to 210K one sees considerable sharpening of the peaks, with the second peak in the OW-OW function becoming notably more pronounced and moving to larger $r$. This behaviour closely resembles what is predicted to occur for bulk water when taken to low density \cite{soper2000structures} and is also seen when low-density amorphous ice (LDA) is formed \cite{bosio1986x,bizid1987structure,finney2002structures}. This second peak is traditionally adopted as an indicator of the degree of tetrahedral order in water, since it occurs at the required $\sqrt{8/3}$ radial position compared to the first peak position for tetrahedral order. Hence if this assignment is correct, the degree of tetrahedral order in confined water at low temperature has certainly increased quite markedly. These data of course tell us nothing about the state of that water, whether it remains a liquid, becomes a glass, or is some form of disordered crystal, although the absence of a clear signature in the DSC trace from MCM41 materials with this pore size \cite{jahnert2008melting} implies no change of phase has occurred on cooling. However that same work shows, from proton NMR cryoporometry, there is likely to be a continuous solid-like to liquid-like transition around 218K at this pore diameter, with no clear information on the nature of the solid phase, but evidence for range of relaxation times as you proceed from the surface of the pore to the centre \cite{findenegg2008freezing}.

\subsection{\label{localdensity}Measurement of the local density for confined water}

The data of Fig. \ref{figwaterrdfs}(a) show that the effect of confinement on fluid structure is to place the fluid-fluid autocorrelation function on a negative slope with increasing $r$. When considering an ensemble of pores, as in the present case, the autocorrelation of the fluid within a single pore function must be convoluted with the distribution of pores to give the total fluid-fluid correlation function, but at small $r<R_{sep}$, where $R_{sep}$ is the shortest distance between the surface of one pore and that of a neighbouring pore, the correlations will come mainly from positions within the same pore. In fact quite general arguments \citep{glatter1982small} suggest that the shape of the single pore autocorrelation function at short distances is linear with negative slope, $c(r) = 1-ar+...$, with $a$ a constant related to the dimension of the pore. Indeed for a solid uniform sphere this function is analytic:

\begin{equation}
 c_{\textrm{sphere}}(r)=
  \begin{cases}
   1-\frac{3}{4}\left(\frac{r}{R}\right)+\frac{1}{16}\left(\frac{r}{R}\right)^3 & \textrm{if } r < 2R \\
   0      &  \textrm{if } r \geq 2R
  \end{cases}
\label{spheregr}
\end{equation}
where R is the radius of the sphere. Hence in this case $a=\frac{3}{4R}$. For a cylinder of length $L$ and radius $R$ there does not appear to be an equivalent analytic form, but the $c(r)$ function can be estimated in this case from the observation that the $Q$-dependent form factor for a solid uniform cylinder is:

\begin{equation}
F(Q,R,L)=\frac{\pi R^{2}L}{2}\int_{-1}^{+1}\left(\frac{\sin Q\mu L/2}{Q\mu L/2}\right)^{2}\left[\frac{2J_{1}\left(Q\sqrt{1-\mu^{2}}R\right)}{Q\sqrt{1-\mu^{2}}R}\right]^{2}d\mu
\label{fqcylinder}
\end{equation} 
where $\mu$ is the cosine of the angle between vector \textbf{Q} and the axis of the cylinder and $J_{1}(x)$ is the integer Bessel function of the first kind. To get the radial dependence of the autocorrelation function $F(Q,R,L)$ needs to be averaged over $\mu$ and Fourier transformed to $r$ space. For the case of an infinitely long cylinder, $L\rightarrow\infty$, and radius $R$, the integrand is finite only for values of $\mu\approx 0$, while for the case of a disk of infinite radius, $R\rightarrow\infty$, and thickness $L$, only values of $\mu\approx 1$ are important. In both cases this means the orientational average can be performed analytically, leading to the following expressions for $c(r)$ for an infinitely long uniform cylinder:

\begin{equation}
c_{\textrm{cyl}}(r)= \frac{2R^2}{r}\int_{0}^{\infty}\left[\frac{J_{1}\left(QR\right)}{QR}\right]^{2} \sin Qr dQ
\label{cylgr}
\end{equation}
and for an infinitely wide uniform disk:
\begin{equation}
c_{\textrm{disk}}(r)=\frac{2L}{\pi}\int_{0}^{\infty}\left[\frac{\sin\left(QL/2\right)}{QL/2}\right]^{2} \frac{\sin Qr}{Qr} dQ
\label{diskgr}
\end{equation}

\begin{figure}
\centering
\includegraphics[scale=1]{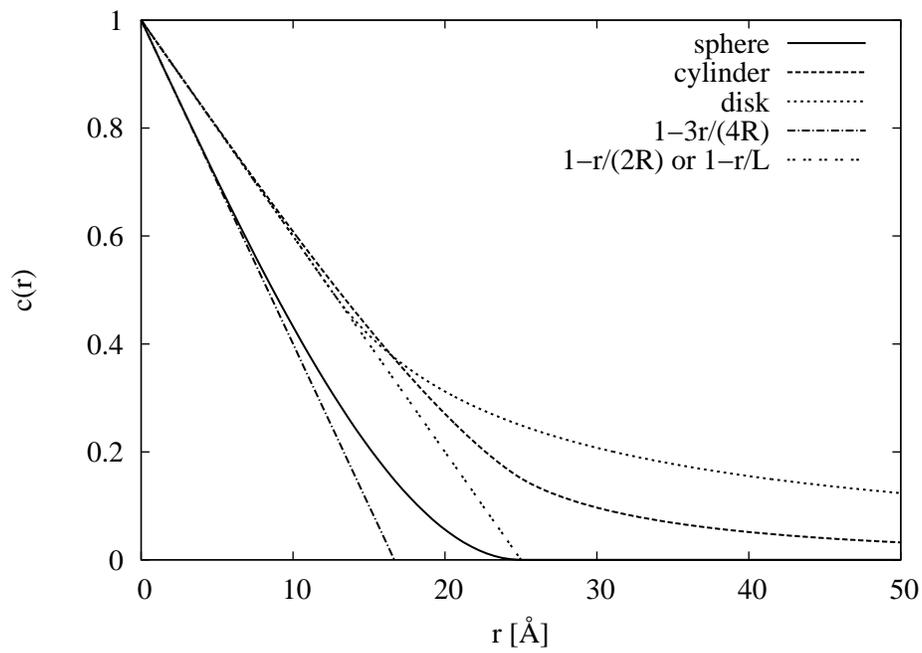}
\caption{Autocorrelation function, $c(r)$ for a sphere of radius $R=12.5$\AA\ (solid), infinitely long cylinder of radius $R=12.5$\AA\ (dashed), and infinitely wide disk of thickness $L=25$\AA\ (dotted). Also shown are the lines corresponding to the sphere at low $r$ (dot-dashed) and the cylinder or disk at low $r$ (intermittent dots).}
\label{figcylspheregrs}
\end{figure}

Equations (\ref{cylgr}) and (\ref{diskgr}) are evaluated numerically for the case where $R=L/2=12.5$\AA\ and are shown together with the spherical case for $R=12.5$\AA\ in Fig. \ref{figcylspheregrs}. It can be seen that in all three cases the autocorrelation does indeed decay initially in a linear fashion with $r$. For the sphere, $a=\frac{3}{4R}$ as expected, while for the infinitely long cylinder and infinitely wide disk, $a=\frac{1}{2R}=\frac{1}{L}$ respectively. Hence the gradient of this initial slope of the autocorrelation function is a measure of the characteristic confinement length, $L_c$, of the confining medium. For cylinders and disks this confinement length will be defined as  $L_c = 1/a$. At larger $r$ the autocorrelation for the isolated sphere drops to zero as anticipated, but for the cylinder and disk remains finite at large distances due to the infinite extent of these objects.

These observations suggest a simple way to measure the local density in the pore. Using the functional form $g_{\alpha\beta}(r') = 1-ar'+...$ in (\ref{nofr}), the number of atoms out to a specified distance, $r$, in the uniform fluid is defined as
\begin{equation}
N_{\alpha\beta}\left(r\right)=\frac{4}{3}\pi\rho_{\beta}^{(l)} r^3 (1-\frac{3ar}{4}+ ...)
\end{equation}
where $\rho_{\beta}^{(l)}$ is the \textit{local} density of $\beta$ atoms in the pore. From this the average density of $\beta$ atoms around a given $\alpha$ atom at the origin in the range $r'=0 \rightarrow r$ is given by
\begin{equation}
\rho_{\alpha\beta}\left(r\right)=\frac{N_{\alpha\beta}\left(r\right)}{\frac{4}{3}\pi r^3}=\rho_{\beta}^{(l)}(1-\frac{3ar}{4}+ ...)
\label{avrhoofr0}
\end{equation}

\begin{figure}
\centering
\includegraphics[scale=0.88]{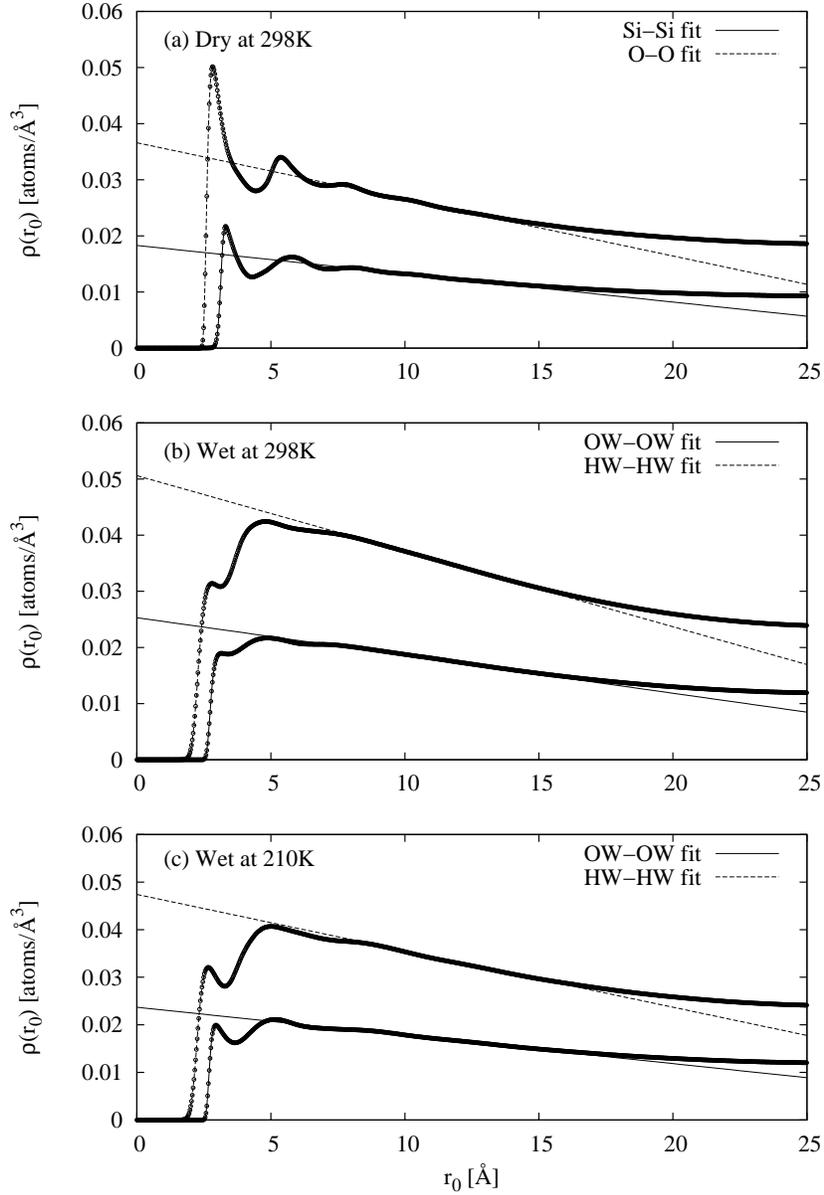}
\caption{Local density calculations (equation (\ref{avrhoofr0}), middle expression) for the Si-Si and O-O radial distribution functions in dry MCM41 (a), and for the OW-OW and HW-HW radial distribution functions in wet MCM41 at 298K (b) and 210K (c). The straight lines defined by equation (\ref{avrhoofr0}), right-hand expression, were fit in the region 5 - 15\AA. Parameters from these fits are given in Table \ref{tablocaldensities}. For O-O and HW-HW it was assumed the local density was twice the Si and OW local densities respectively, and the confining dimensions for O and HW were assumed to be the same as for Si and OW respectively.}
\label{figlocaldensitycalc}
\end{figure}

In the limit $r \Rightarrow 0$, $\rho_{\alpha\beta}(0) = \rho_{\beta}^{(l)}$, so estimating $\rho_{\alpha\beta}\left(r\right)$ from the real $g(r)$s, such as in Fig. \ref{figwaterrdfs}(a), using (\ref{nofr}), and linearly extrapolating these to $r=0$, using the right-hand side of (\ref{avrhoofr0}), gives us simultaneously a measure of the local density, $\rho_{\beta}^{(l)}$, and the approximate size of the confining medium, $L_{c}=1/a$, via the gradient coefficient $a$. These calculations are shown in Fig. \ref{figlocaldensitycalc} for the Si-Si and O-O distributions in the dry MCM41, and the OW-OW and HW-HW distributions in the wet MCM41. The parameters derived from the linear fits are given in Table \ref{tablocaldensities}

\begin{table}
\begin{tabular}{|c|c|cccc|}
\hline 
Atom & Distibutions & $\rho_{\beta}^{(l)}$ & L$_{c}$ & Core & Bulk\\ 
• & fitted & [at./\AA $^{3}$] & [\AA] & [at./\AA $^{3}$] & [at./\AA $^{3}$]\\ 
\hline 
\small{Si} & \small{Si-Si,O-O} & 0.0183(2) & 27.2(1) & - & 0.0221 \\  
\small{OW} (298K) & \small{OW-OW,HW-HW} & 0.0253(2) & 28.2(1) & 0.031(1) & 0.0334 \\ 
\small{OW} (210K) & \small{OW-OW,HW-HW} & 0.0237(2) & 30.0(1) & 0.027(1) & (0.0307)\\ 
\hline 
\end{tabular}
\caption{\label{tablocaldensities}Estimated local densities for Si and OW atoms in MCM41 based on the linear extrapolation of $\rho_{\alpha\beta}\left(r\right)$ to $r=0$ using the right-hand side of (\ref{avrhoofr0}). The corresponding confinement lengths, $L_c=1/a$, where the gradient of the fit lines is given as $\frac{3a}{4}$ were determined at the same time. Also shown are the corresponding core densities (for OW) as determined in section \ref{densityprofile} from Fig. \ref{figwetprofiles}, and the bulk atomic number densities for the same atoms. For fitting the O-O and HW-HW distributions it is assumed the local density of the O and HW atoms is exactly twice that of the corresponding Si and OW atoms. The uncertainties are measured by determining the amount of variation in the respective quantity which would be needed to make a discernible change to the fit. The bracketed value for OW at 210K corresponds to the value for Ice Ih at 273K.}
\end{table}

Several comments can be made here. Firstly it is observed in Fig. \ref{figlocaldensitycalc} that there is indeed a nearly linear region in the local density function, as anticipated by equation (\ref{avrhoofr0}). Secondly from Table \ref{tablocaldensities} the local densities are significantly lower than their bulk counterparts. This arises quite simply because a fraction of the water has penetrated the silica substrate. Equally some of the silica is inside the nominal 12.5\AA\ radius of the pore, Fig. \ref{figdrymcm41profiles}. In both cases this means the local density, when averaged over the unit cell volume, will fall below the value it would have if the water were confined strictly inside the pore, and silica strictly outside of it. Thirdly we note that the water oxygen local density falls by approximately 9\%\ when going from 298K to 210K, a trend which is mirrored by the known densities of water at 298K and ice at 273K. This fall happens even though the simulation box at each temperature contains the same number of water molecules. At the same time the confinement length $L_{c}$ grows slightly from 28.2\AA\ to 30.0\AA as the temperature is lowered. The implication is therefore that confined water has pushed slightly further into the confining medium on reducing the temperature, causing the lower density. The reader will readily appreciate that this is similar to the (relative) fall in density observed by S-H Chen and coworkers, who used small angle neutron scattering to study the intensity of the (100) Bragg peak from D$_2$O absorbed in MCM41 \cite{chen2007:4,zhang2011density,kamitakahara2012temperature}. In that case the assumption, based on previous dynamic data \cite{savedrecs2005:1}, is that confined water remains a liquid at 210K, but here there is no evidence for or against that supposition.

\subsection{\label{densityprofile}Density profile across the pore - the core density}

\begin{figure}
\centering
\includegraphics[scale=1]{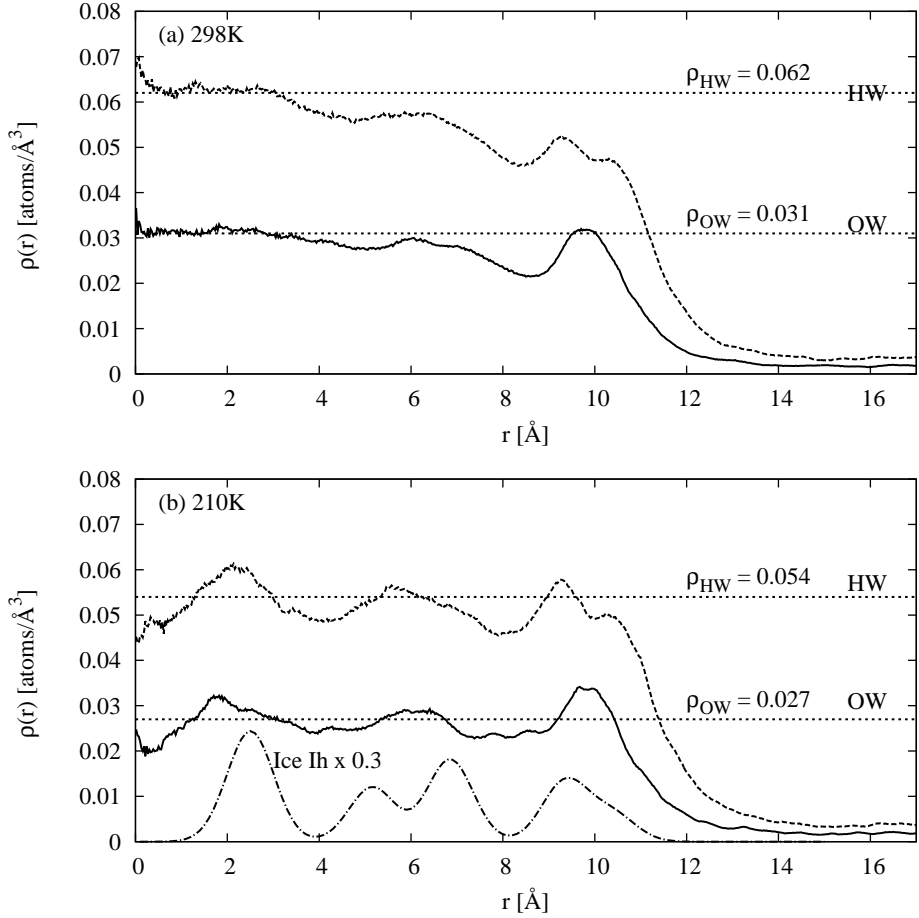}
\caption{Water density profiles for OW (solid) and HW (dashed) atoms as a function of distance from the centre of the pore, at 298K (a) and 210K (b). Also shown (dotted) is a horizontal line corresponding to the average density in the range $r = 0 - 4$\AA at each temperature. This density is called the ``core'' density. The dot-dash line shows a simulation of ice Ih with the ice $c$-axis aligned with the cylindrical pore axis and the centre of hexagon at the centre of the pore (see text).}
\label{figwetprofiles}
\end{figure}

In order to characterise further the nature of the water density when confined in MCM41, it is informative to show the density profile as a function of distance from the centre of the pore, Fig. \ref{figwetprofiles}.

At 298K the water oxygen density profile is relatively smooth as a function of distance from the pore, starting from a value close to the bulk value at the centre of the pore, then declining slowly to about 8.5\AA\ from the pore centre, followed by a pronounced peak which forms the main layer of water near the surface of the pore. Beyond this the density profile decays rapidly as it overlaps with the substrate region. Hence this density profile corresponds rather closely to what has been stated previously \cite{soper2012density} in terms of there being three regions in confined water, namely core, interfacial and overlap. Note however that even in the core region, 0-4\AA, the water oxygen number density, 0.031 atoms/\AA$^{3}$, is significantly lower than that found in bulk water (0.0334 atoms/\AA$^{3}$) at the same temperature - see Table \ref{tablocaldensities}.

At 210K the density profile changes, becoming more uneven and with a more pronounced interfacial peak. The overall core density falls however, and there is less evidence the density would approach even the bulk value for ice (0.0303 atoms/\AA$^{3}$) at the centre of the pore. The fact that the structure can apparently support significant density variations as a function of radius from the pore centre would argue against this being a true liquid at this temperature: certainly at least the diffusion would have had to slow significantly compared to at 298K where the density variation as a function of distance is much smoother. Note that the simulations at both temperatures were run for the same number of Monte Carlo steps, and there was no sign of the density profile evolving or becoming smoother as the simulations were run longer. These results are therefore in agreement with those presented in \cite{mancinelli2010controversial} using an independent analysis. In \cite{mancinelli2010controversial} there was indeed a marked increase in unevenness of the density profile in the core region on cooling below 210K compared to 298K, even though the adopted size of the pores was smaller in that work.

To further illustrate the point that the core may be solid at 210K, the dot-dash line in Fig. \ref{figwetprofiles} shows a simulation of the ice Ih structure, based on the lattice constants given in \cite{rottger1994lattice} at this temperature, in which the ice crystallographic $c$-axis is set parallel to the cylinder axis and the centre of an ice hexagon is placed at the centre of the pore. The ice distribution has been broadened using a Gaussian of width 0.5\AA\ to simulate the likely disorder that would occur in such a situation. Obviously the correspondence of the peaks between ice Ih and the simulated density profile may be purely coincidental, but the comparison does illustrate that a disordered crystalline or solid structure would not be incompatible with this degree of confinement. The particle size broadening incurred by the high degree of confinement would probably preclude direct observation of the Bragg peaks associated with this structure in the scattering pattern, making it appear disordered \cite{mancinelli2010effect}. 

\begin{figure}
\centering
\begin{tabular}{c}
(a) 298K\\
\includegraphics[scale=0.2]{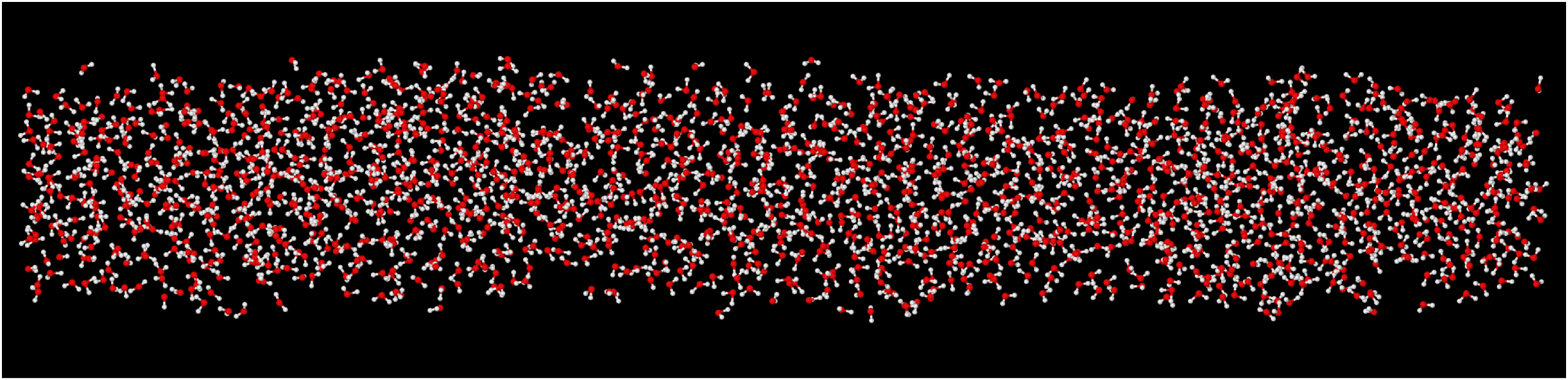} \\
(b) 210K\\
\includegraphics[scale=0.2]{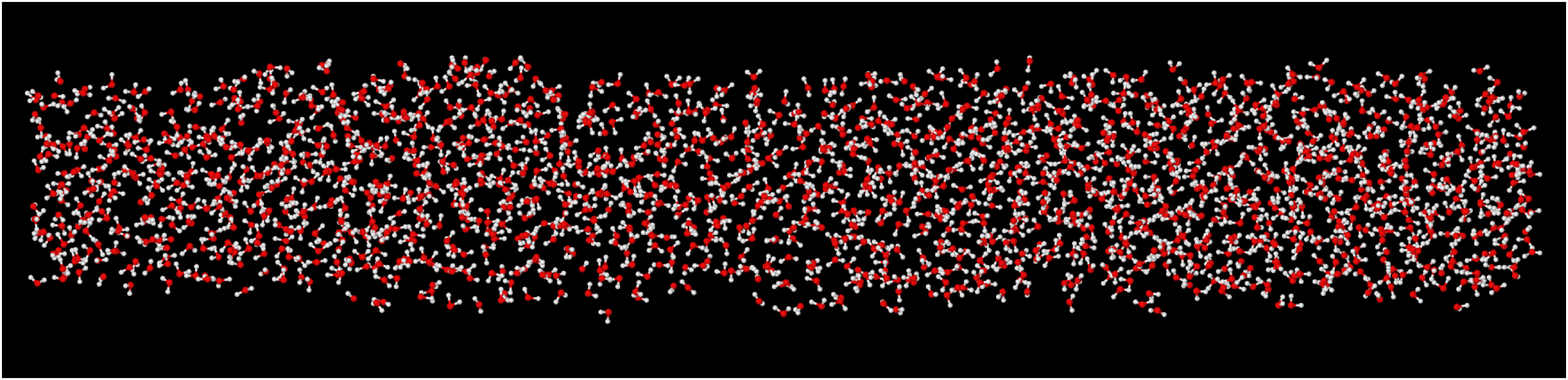} \\
\end{tabular}
\caption{\label{figplotpores}(Colour online) Plots of the EPSR simulation of water in MCM41 for one of cylindrical pores at 298K (a) and 210K (b). There is little sign of obvious density fluctuations along the pore at either temperature.}
\end{figure}

A related question is the extent to which density fluctuations occur \textit{along} the pores, \cite{mancinelli2010controversial}. To illustrate this, Figure \ref{figplotpores} shows the water in one of the pores for each of the temperatures 298K and 210K. As noted there is little sign of obvious density fluctuations along the pore, at least on the timescale of the current simulations.

\subsection{\label{qdefinition}Structural changes with temperature - tetrahedrality parameter}

In order to characterise tetrahedral order in water, Errington and Debenedetti \citep{errington2001relationship} introduce the $q$ order parameter based on the angle between triplets of neighbouring water oxygen atoms:-

\begin{equation}
q=1-\frac{3}{8}\sum_{j=1}^{3}\sum_{k=j+1}^{4}\left(\cos\theta_{jk}+\frac{1}{3}\right)^2
\end{equation}
where the sum over $j$ and $k$ covers the six triplets of angles which involve a given water oxygen atom and its 4 nearest neighbours. This value is averaged over all the water molecules in the simulation box. For water in confinement (or concentrated solution), where the local density may be significantly lower than in the bulk liquid, the use of the 4 nearest neighbours could cause the value of $q$ to appear small because one or more of those 4 neighbours are outside the nearest neighbour distance. An alternative method of calculating $q$ is to define the expected nearest neighbour distance, based (for example) on the position of the first minimum in the OW-OW radial distribution function, then calculate the distribution of included angles, $N(\theta)$, which involve triplets of water molecules, at least two pairs of which are at or below this cut-off distance. The included angle is that associated with the common water oxygen atom. (If all three water molecules are within this cut-off distance, then that counts as three triplets.) To give the \textit{density} of triplet angles the $\sin\theta$ distribution that would occur with completely random atomic positions has to be divided out: $P(\theta)={N(\theta)}/{\sin\theta}$. Using this density of triplet angles, $q$ can be redefined for arbitrary concentration or degree of confinement:

\begin{equation}
q=1-\frac{9}{4}\frac{\int_0^{\pi}P(\theta)\left(\cos\theta+\frac{1}{3}\right)^2\sin\theta d\theta}{\int_0^{\pi}P(\theta)\sin\theta d\theta}
\end{equation}
where the factor of $\frac{9}{4}$ is required to ensure $q$ goes from 0 ($P(\theta)= \mathrm{constant})$ to 1 ($P(\theta)=\delta(\cos\theta-1/3)$). In this case, instead of averaging $q$, it is $P(\theta)$ that is averaged over the simulation box and over molecular configurations. This average distribution, $\langle P(\theta)\rangle$, is then used to calculate $q$.

\begin{figure}
\centering
\includegraphics[scale=1]{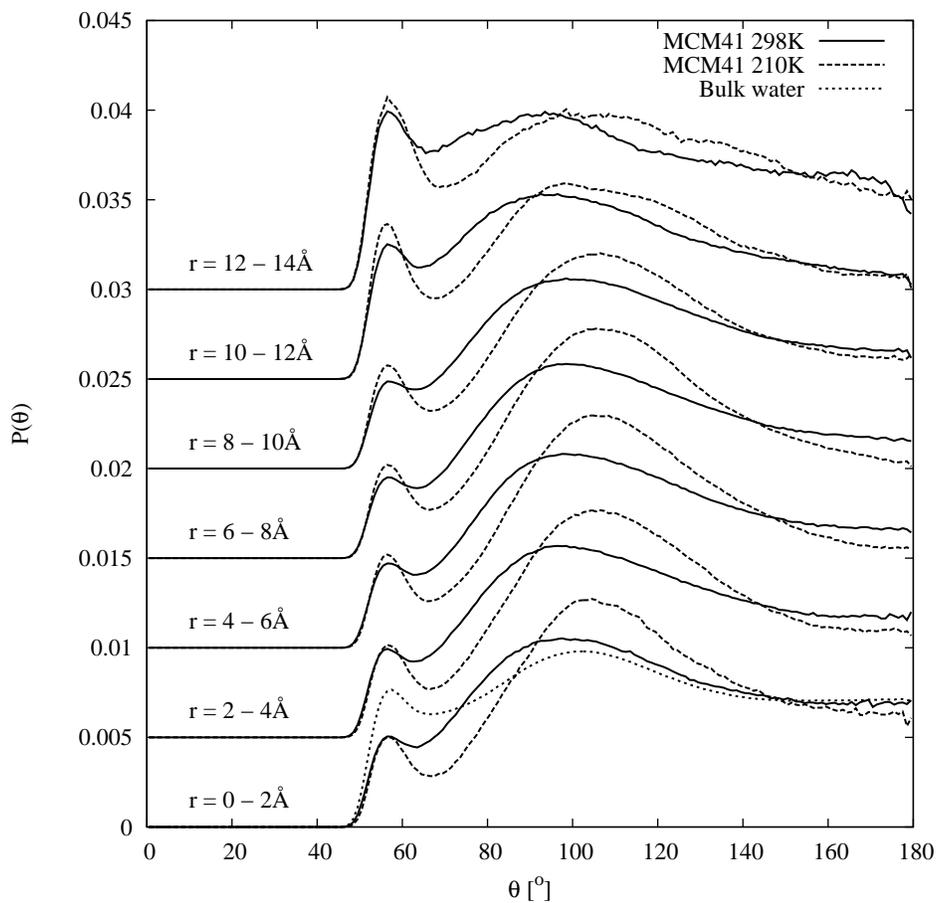}
\caption{OW-OW-OW included angle distribution, $P(\theta)$, for the specified distance ranges from the centre of the pore at 298K (solid) and 210K(dashed). To define each neighbour of the triangle the maximum OW-OW distance was set 3.24\AA. The dotted line shows the same distribution for bulk water at 298K, determined from EPSR simulation of the merged diffraction data presented in \cite{soper2013radial} with the same reference potential as used here.}
\label{figwettriangles}
\end{figure}

\begin{figure}
\centering
\includegraphics[scale=1]{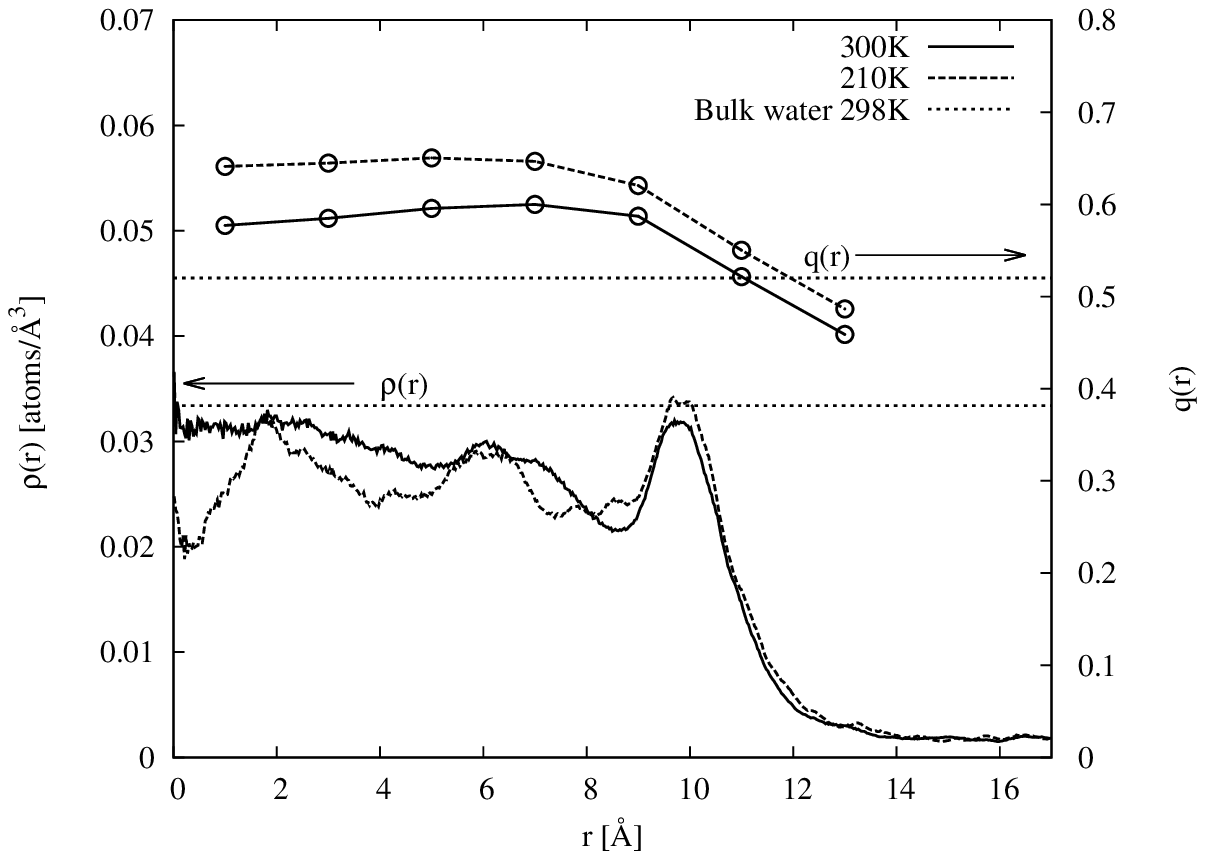}
\caption{Variation of absorbed water density as a function of distance from the centre of the MCM41 cylindrical pore (bottom plots, left-hand scale). Variation of $q$ over the same distance range (top, right-hand scale). The results at 298K are shown as the solid lines, and those at 210K by the dashed lines. The dotted lines show the respective values for bulk water.}
\label{figqvariation}
\end{figure}

In the present case, the OW-OW cut-off distance was set to 3.24\AA, that being close to the position of the first minimum in the OW-OW radial distribution function of confined water, Fig. \ref{figwaterrdfs}. As a guide, the value of $q$ obtained from EPSR simulated water with this cut-off distance, using the SPCE reference potential and the latest combined x-ray and neutron scattering data \citep{soper2013radial} is 0.52. The distribution of included angles at various distance ranges is shown in Fig. \ref{figwettriangles}, while the corresponding values of $q$ at the same distance ranges are shown in Fig. \ref{figqvariation}.

The distribution of included angles for ambient water typically consists of a broad hump near 100$^o$ corresponding to loosely tetrahedrally bonded arrangements of triplets, plus a smaller peak or shoulder near or below 60$^o$ corresponding to triplets for which at least one pair the hydrogen hydrogen bond is heavily distorted or broken. The former peak will be referred to as the ``tetrahedral'' peak  while the latter as the ``interstitial'' peak. The shape of this distribution is a sensitive indicator of the impact of solutes on water structure \cite{mountain2004importance}, although the detailed shape can also depend on the intermolecular potential used in the simulation. For the present case at 298K it can be seen that this distribution follows the observed pattern in bulk water, but with a reduced interstitial peak, and slightly enhanced tetrahedral peak. At 210K the tetrahedral peak becomes more pronounced and moves closer to the ideal tetrahedral angle of 109.47$^o$. However near the pore surface the distribution becomes heavily distorted at both temperatures, with increased interstitial peak, signalling a breakdown of the normal water structure in this region. 

Corresponding to these changes it can be seen, Fig. \ref{figqvariation}, that in the centre of the pore, $q$ is significantly \textit{above} the value for bulk water, and this value increases markedly when the temperature is lowered. In the interfacial and overlap regions however the value at both temperatures falls below the ambient bulk water value, signalling a collapse of tetrahedral structure near the surface. Based on these results therefore it would appear that in the centre of the pore, confined water is actually \textit{more} tetrahedral than in the bulk, even at 298K, a trend that might arise from the overall lower density of core water compared to bulk water.

\section{\label{discussion}Discussion}

The foregoing account draws heavily from previous RMC or EPSR treatments of water in confinement, \cite{iiyama2006direct,thompson2007three,mancinelli2009multiscale,mancinelli2010effect}. Nonetheless there are some important changes. In particular the present work uses a pore size which is more consistent with the known amount of water absorbed in the pores and also the behaviour of the (100) Bragg diffraction peak, wet and dry, as a function of hydrogen isotope \cite{soper2012density}. In addition the simulated structure is fit to \textit{both} the hexagonal Bragg peak intensities \textit{and} the wider $Q$ scattering pattern, something that has not been attempted previously. Although still not giving a completely unambiguous view of the structure, the extra constraints imposed by including the Bragg intensities make it difficult to see how the final conclusions could be radically different from those presented here. Given the somewhat contentious nature of what is known about water in confinement (a recent review \cite{bertrand2013deeply} seems to tell only part of the story) it is important to establish the degree of certainty of various statements about water in confinement.

\subsection{\label{density}The density of confined water}

For a bulk fluid, or for a fluid confined by well-defined walls, density can be defined rather precisely as the amount of material or number of atoms in a known volume. When the walls become soft and fluid can penetrate the wall to a greater or lesser extent, as in many real instances, this precise definition loses its meaning and our proposal here is that we should instead talk about the ``local'' and ``core'' densities of the fluid in such cases. For the case reported here both the local and core densities of confined water, Table \ref{tablocaldensities} fall by about 7\%\ when confined water is cooled from 298K to 210K, an amount which is closely similar to the change in density when water freezes, or between ambient water and low density amorphous ice. However the \textit{absolute} values of these densities at 298K are already significantly lower than bulk water density, by 6\%\ for the core density at this temperature. So what is the evidence for these assertions?

\begin{figure}
\centering
\includegraphics[scale=1]{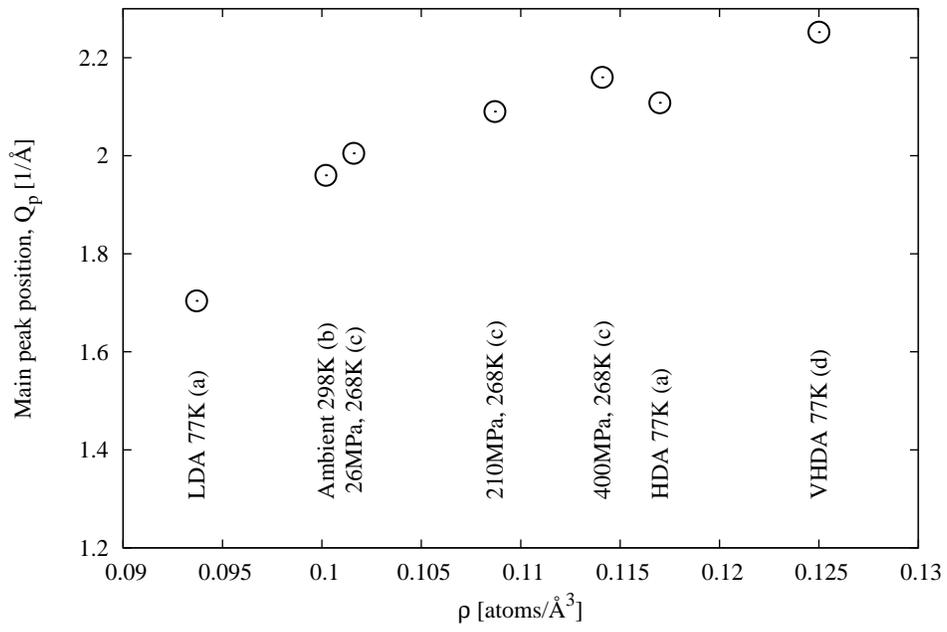}
\caption{Position of the main peak in the diffraction pattern for water and amorphous ice (D$_{2}$O) as a function atomic number density. The data are taken from (a) \citep{finney2002structures}, (b) \citep{soper2013radial}, (c) \citep{soper2000structures} and (d) \cite{finney2002structure}.}
\label{figD2Opeakpositions}
\end{figure}

One piece of evidence is the position of the main D2O diffraction peak for water in confinement. In the current work (see Figs. \ref{figwetmcm41fits} and \ref{figwetmcm41fits210K}) the main D2O diffraction peak occurs at $\sim$1.86\AA$^{-1}$ at 298K and $\sim$1.72\AA$^{-1}$ at 210K: both values are significantly below the value for ambient water at 298K, $\sim$1.95\AA$^{-1}$. A similar trend has been observed in other work on water in confinement, \cite{yoshida2008thermodynamic,futamura2012negative}. Figure \ref{figD2Opeakpositions} compares the first peak position for liquid water and amorphous ice (D$_2$O) at different densities. Obviously one cannot take the first peak position as a direct measure of density, since it contains contributions from all of the OW-OW, OW-HW and HW-HW terms, nonetheless there is a very clear trend here: as the density falls, whatever the temperature or pressure, the main peak moves to lower $Q$. Hence the fact that in confined water the main diffraction peak is consistently below its position in the bulk liquid \cite{kamitakahara2012temperature} is already an indication that confined water is at lower density than the bulk. The contention by Liu et al. \citep{liu2007observation} that the density of confined water is continuous with that of supercooled bulk water at the same temperature is not supported by these data. 

Another piece of evidence is the height of the (100) Bragg peak. As seen in Fig. \ref{figwetmcm41fits} there is a factor of $\sim$ 4.6 reduction in the height of this peak when substituting  H$_{2}$O for D$_{2}$O at 298K. If, for the time being, we make the assumption, that the density profile is flat and sharply defined at the edges of the pore, then the change in height of the peak relates simply to the square of the difference in scattering length density between water and silica \citep{liu2007observation}. The total scattering length of a D$_{2}$O molecule is 19.14fm, while for H$_{2}$O it is -1.68fm and for the SiO$_{2}$ unit it is 15.76fm. Assuming a number density for the silica units the same as bulk silica, 0.0221/\AA$^{3}$, and making allowance for the silanol groups (the hydrogen atoms of which will exchange readily with the water hydrogens), then the scattering length densities of the D and H substrates are $\rho_{s}^{(D)}=0.360$fm/\AA$^{3}$ and $\rho_{s}^{(H)}=0.290$fm/\AA$^{3}$. The (unknown) molecular number density of the water is $\rho_{w}$ so the ratio of Bragg intensities for light water to heavy water is given by:

\begin{equation}
\frac{I_{H_2O}}{I_{D_2O}}=\frac{\left(1.68\rho_w+\rho_{s}^{(H)}\right)^{2}}{\left(19.14\rho_w-\rho_{s}^{(D)}\right)^{2}}
\label{eqintensityratio}
\end{equation} 

\begin{figure}
\centering
\includegraphics[scale=1]{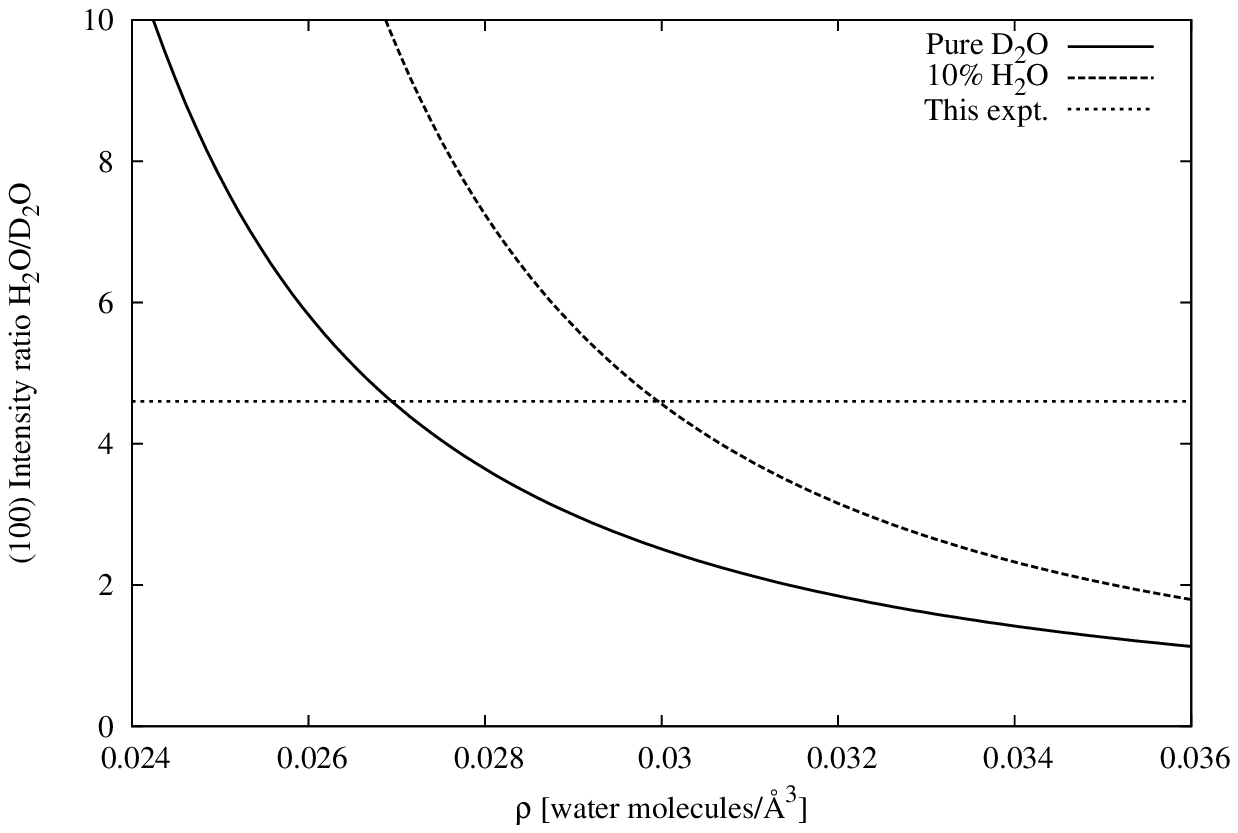}
\caption{Predicted (100) Bragg peak intensity ratio (solid), H$_{2}$O/ D$_{2}$O, equation (\ref{eqintensityratio}) as a function of water density inside the pore, assuming a uniform density profile and sharp edges at the pore wall. The density of the silica is taken to be that of bulk silica. Also shown is the case where the D$_{2}$O is contaminated with 10 mole\%\ H$_{2}$O (dashed). The dotted line shows the ratio found in the current experiment.}
\label{figintensityratios}
\end{figure}

This ratio is plotted in Fig. \ref{figintensityratios}, but note that this approach makes the highly simplifying assumptions that the water is distributed uniformly across the pore and makes no penetration into the wall of the pore. It can be seen readily that to obtain a peak ratio as high as 4.6, the density of the water has to be significantly below ambient (0.0334 molecules/\AA$^{3}$). This conclusion is maintained even in the event that the heavy water is contaminated with some light water, as assumed in the current data analysis and simulations. It can also be seen in this figure that to get the bulk density of water in the pore, the peak ratio would have to be $\sim$3 or below. The conclusions from this figure are borne out by the present simulations: if too much water is added to the (simulated) pore, the intensity of the (100) peak in the simulated D$_{2}$O spectrum becomes too large compared to the scattering data. However, as was seen in \cite{soper2012density}, if the density profile inside the pore is non-uniform (which it almost certainly is), then other factors come into play to determine the peak ratio, so this simple analysis doesn't work.

In contrast, Liu et al. performed a similar analysis to the above, \cite{liu2008density}, but came to the opposite conclusion, namely that, for the fully hydrated pore, confined water has a density 8\%\ \textit{higher} than the bulk density. This is in contrast to the same authors' work of 2007 \cite{liu2007observation} where it is assumed the confined density is the same as the bulk density. However the observed H$_{2}$O/ D$_{2}$O peak ratio is closer to 3 in that work, compared to the present 4.6, which would indeed apparently imply a higher density than reported here. Even with this lower peak ratio, and assuming the heavy water has no contamination with light water, the solid line in Fig. \ref{figintensityratios} would still imply a density inside the pore lower than the bulk. One detail is different, namely the assumed scattering length density of the substrate, which here is based on the density of bulk silica, but in \cite{liu2008density} is measured to be $0.4006$fm/\AA$^{3}$. Where such a large value arises from is not clear. Moreover of the two computer simulations which are cited to support the view of an increased density of confined water in MCM41, one \cite{merzel2002first} is to do with water at the surface of lysozyme and therefore really has no relevance to the case of water inside the highly hydrophilic and cylindrical pore created by amorphous silica. The other \cite{lee1994comparison} is to do with water against a crystalline silica slit pore of much larger dimensions than those being considered here. Enhanced density occurs in the well-defined surface water layer as expected (it is of course preceded by a region of zero density immediately adjacent to the crystalline wall), but across the pore as whole there is no evidence for an 8\%\ density increase.

In addition there are other differences compared to the present work. that need to be considered. In the data of \cite{liu2008density} only the (100) diffraction peak is shown so it is impossible to guage what is the local atomic structure in this material. Hence all the detail of the density profile of both the water and the silica substrate is lost in that work. The (100) diffraction peak occurs at $\sim$0.192\AA$^{-1}$ compared to 0.215\AA$^{-1}$ in the present work, which implies a lattice constant of 37.79\AA - much larger than the 33.1\AA\ of the present data. It is also stated that the amount of D$_2$O absorbed is 0.5g/g  of MCM41, which translates to 0.45g/g for the corresponding amount of H$_2$O. Solving equation (11) of \cite{soper2012density} gives a pore diameter of 28\AA, which is significantly larger than the stated 19\AA. This discrepancy serves to reinforce the view of this author that before making any statements about density and its trend with temperature, a full analysis of the entire scattering pattern is required: increasing the diameter of the pores by 9\AA\ can have a dramatic effect on the calculated Bragg peak intensities, water density profile and water structure. It also borders on the region where core water in confinement crystallises when cooled \cite{jahnert2008melting}. The analysis of \cite{liu2008density} makes no reference to the \textit{shape} of the density profile across the pore, even though that profile can be critical to determining the relative heights of Bragg peaks. Furthermore that same analysis assumes the silica density profile is structureless, which it clearly is not, Fig. \ref{figdrymcm41profiles}.

Based on the evidence presented here, therefore, the conclusion seems unavoidable that the density of water confined in MCM41 is significantly lower than in the bulk liquid at the same temperature. Exactly how much below the bulk density is debatable, but a decrease of order 10\%\ in the core region of the pore seems consistent with several pieces of evidence. Independent evidence \cite{jahnert2008melting} also supports the view that water confined in MCM41 has lower density than the bulk. The density profile across the pore is very far from uniform, with a marked density increase in the interfacial region, and in addition the confined water has a significant overlap with the silica substrate. Most likely the density profile changes when lowering the temperature to 210K, becoming more structured, but also pushing towards the edges of the pore. This in turn gives rise to a lowering of both the local and core densities by an amount consistent with what is stated elsewhere \cite{liu2007observation}. This work emphasizes the importance of obtaining scattering patterns over the full $Q$ range, and not relying simply on one region to make conclusions about the nature of the confined medium.

\subsection{\label{poreradiusdisc}Pore radius}

The dimension of the pores is also a topic which has led to marked differences in reported results. Besides the analysis presented in \cite{soper2012density}, independent corroborating evidence suggests that gas adsorption measurements on their own can seriously underestimate the pore diameter, \citep{jahnert2008melting}, by as much as 6\AA. Indeed the methods proposed in \cite{mancinelli2009multiscale,soper2012density} closely follow the x-ray analysis of \cite{jaroniec2006improvement}, which led to the conclusion that pore diameters estimated by gas adsorption needed to be revised upwards. An earlier study comparing pore diameters obtained by NMR cryoporometry with those obtained from neutron scattering also suggested larger diameters than traditionally supposed, \cite{webber2001evaluation}. The adoption of the nominal pore diameter of 25\AA\ (radius 12.5\AA) in the present work, which is both consistent with observed amount of water absorbed in the pore, and with the trends of the Bragg peaks with water absorption and isotope changes, seems eminently reasonable in the light of those studies. 

\subsection{\label{waterstructureconfinement}The nature of water structure in confinement}

The data given in Figs. \ref{figwaterrdfs}(a) and \ref{figwettriangles} contain two elements to be highlighted, namely the fact that under ambient conditions, water contained in MCM41 is more tetrahedral than bulk water, and is almost certainly at lower density, even in the core region of the pore. Also that on cooling to 210K there is a marked change in structure of confined water with increased tetrahedral order in the liquid compared to bulk water, Fig. \ref{figqvariation}, and even lower both local and core densities, \ref{figlocaldensitycalc}. A number studies attest to the fact that this structural transition is accompanied by a dynamic transition \cite{faraone2004fragile,mallamace2006fragile,liu2006quasielastic,mallamace2007evidence,chen2009evidence} from fragile liquid to strong liquid. However others are more cautious, pointing out that even when the core of the pore is solid there is NMR evidence for one or two layers of disordered water near the surface which remain mobile to low temperature \citep{schreiber2001melting,findenegg2008freezing,jahnert2008melting}. Even more recent and independent work using calorimetry also argues against the transition that occurs in confined water on cooling to 210K being a either a glass-liquid or liquid-liquid transition \cite{tombari2012specific,tombari2013state}. The present work cannot comment on this dynamic interpretation, except there does indeed appear to be a more disordered layer of water near the surface, and that at 210K the density profile across the pore develops a residual variation with radius, Fig. \ref{figwetprofiles}, which is not dissimilar to what might appear if disordered ice Ih were present and which would imply a more solid-like than liquid-like structure.

\section{\label{conclusion}Concluding remarks}

The observation in the present work of a markedly lower density for water confined in MCM41 silica pores, with an associated increased tetrahedral ordering, compared to bulk water, raises an intriguing possibility concerning the nature of water at this high degree of confinement. Recently there has been renewed interest in the earlier work of Angell and coworkers \cite{zheng1991liquids} on the properties of water under negative pressure, i.e. under tension, when confined in quartz microcavities \cite{caupin2012exploring,azouzi2012coherent,debenedetti2012physics}. In particular it is now believed that at a tension of $\sim$-130MPa the temperature of maximum density occurs near 300K \cite{azouzi2012coherent}, with a maximum density of 922.8kg/m$^{3}$. What relevance to confined water does this result have? Well firstly in both cases the water is contained in a form of silica, SiO$_{2}$. Secondly the present data \cite{mancinelli2009multiscale} relate to a degree of water confinement where there is little or no water \textit{outside} the pores, so the water inside the pore will be in equilibrium with its vapour outside the pore. However water obviously must wet the inside surface of the pore, giving a contact angle close to zero, and so inducing a concave meniscus where vapour meets liquid. If we assume the surface tension of water at this length scale is unchanged from its macroscopic value, 0.072N/m, and given the pore radius of $\sim$12.5\AA\ (1.25nm), the pressure \textit{inside} the pore would have to be negative to the value of $\sim$-115MPa. While there is obviously some uncertainty about this estimate, it does seem consistent with both observations of reduced density in the pore and the increased tetrahedrality which occurs as a result. The same negative pressure and reduced density would also explain why the first main diffraction peak is invariably shifted to lower $Q$ values in confined water compared to bulk, Fig. \ref{figwetmcm41fits} here, Figure 3 of \cite{yoshida2008thermodynamic} and Fig. 3 of \cite{kamitakahara2012temperature}. Indeed the water (oxygen) density obtained in the current simulations at the centre of the pore ($r$=0,  Fig. \ref{figwetprofiles}) is $\sim$0.031 molecules/\AA$^{3}$ at 298K, which translates to a macroscopic density of 927 kg/m$^{3}$. This is consistent with the density of water under negative pressure of -130MPa as reported above. 

If this scenario is correct, are there consequences for what might happen when this confined water is cooled to 210K? According to the Speedy stability limit conjecture, \cite{speedy1982stability}, doing so would inevitably mean we had gone through the low-temperature arm of the stability limit spinodal, causing a crystallisation or solidification event, which would reduce the surface tension at the pore entrance to zero and the pressure would return to ambient. Alternatively if the 2nd critical point scenario is to be believed \cite{poole1992phase}, then we could either have gone through the liquid-liquid transition itself (if the 2nd critical point pressure is negative) or through the line of density fluctuation maxima above this point (sometimes called the ``Widom'' line) if the 2nd critical point is at positive pressure. In the case of a liquid-liquid transition the high degree of confinement would probably prevent a sharp transition being observed. Unfortunately the present data do not tell us unambiguously whether the final state is a disordered crystal, a glass or a liquid, but even if it were a glass or crystal there is a strong likelihood that water molecules near the pore surface would remain mobile \cite{jahnert2008melting}, giving the dynamical signature of a liquid. The only hint here, and it is subject to debate, is the observation of residual density fluctuations across the pore at 210K, Fig. \ref{figwetprofiles}, which would be more likely if the core water was immobile. Another consequence of water in the pore being under significant tension is that a liquid-liquid transition could not occur on cooling, since the confined water would already be in its low density form.

Naturally the fits to the scattering data presented here are not perfect - indeed there remain uncertainties as to the precise amount of water absorbed in the pore and the extent to which the D$_{2}$O data might have been contaminated with H$_{2}$O - which means some details of the structure remain to be understood. Equally the conclusions concerning the density profile differ from those found previously \cite{mancinelli2009multiscale,mancinelli2010controversial,soper2012density} using the same data. However in \cite{mancinelli2009multiscale,mancinelli2010controversial} the assumed pore size was significantly smaller than that adopted here, and there was no attempt to fit the hexagonal lattice Bragg peaks. The Bragg peaks were the \textit{only} features that were fitted in \cite{soper2012density} and the density profile was assumed to have a very simple form. Here an attempt has been made to build an atomistic structural model which is consistent with \textit{all} the scattering data over a wide range of $Q$. Even with these caveats, the main findings of this work, namely a lower than bulk density in the pore and increased tetrahedrality that occurs both on confinement and on cooling to 210K seem robust, given all the evidence presented: changing the amount of material in the simulation box, or adopting a different degree of H$_{2}$O contamination does not materially affect these primary conclusions. Most important is the fact that there is now in a place a methodology for studying the microscopic arrangement of atoms and molecules near surfaces which is consistent with a broad range of observable data. Unless that wide range of data is utilised, the conclusions are likely to be misleading.

\section*{\label{acknowledgements}Acknowledgements}

The author wishes to thank Rosaria Mancinelli, Fabio Bruni, and Maria Antonietta Ricci for unrestricted access to their scattering data on water in MCM41. 





\bibliographystyle{model1a-num-names}







\end{document}